\documentclass[12pt,preprint]{aastex}
\begin{document}

\parindent=1.0cm

\title{And the Rest: The Stellar Archeological Record of M82 Outside the Central Starburst\altaffilmark{1}}

\author{T. J. Davidge}

\affil{Herzberg Institute of Astrophysics,
\\National Research Council of Canada, 5071 West Saanich Road,
\\Victoria, BC Canada V9E 2E7\\ {\it email: tim.davidge@nrc.ca}}

\altaffiltext{1}{Based on observations obtained with the Mehaprime/MegaCam, 
a joint project of the CFHT and CEA/DAPNIA, at the Canada-France-Hawaii Telescope, 
which is operated by the National Research Council of Canada, the Institut National 
des Sciences de l'Univers of the Centre National de la Recherche Scientifique of France, 
and the University of Hawaii.}

\begin{abstract}

	Deep images obtained with MegaCam and WIRCam on the 
Canada-France-Hawaii Telescope (CFHT) are used to probe the 
stellar content outside of the central star-forming regions of M82.
Stars evolving on the asymptotic giant branch (AGB)
are traced along the major axis out to projected distances of 12 kpc, which corresponds 
to 13 disk scale lengths. The numbers of red supergiants 
(RSGs) and AGB stars normalized to local surface 
brightness (the `specific frequency' -- SF) is constant when R$_{GC} > 4$ kpc, indicating 
that RSGs and AGB stars are well mixed throughout the disk. Moreover, 
the SF of bright AGB stars in the outer disks of M82 and the Sc galaxy NGC 2403 are 
identical, suggesting that the specific star formation rates (SFR) in these galaxies during 
intermediate epochs were similar. This similarity in stellar content, coupled with 
the presence of an extended stellar disk, is consistent with M82 having been a late-type 
disk galaxy prior to interacting with M81. Still, there is a paucity of RSGs in the 
outer disk of M82 when compared with NGC 2403, indicating that the SFR 
in the outer regions of M82 during the past $\sim 0.1$ Gyr 
has declined dramatically with respect to that in isolated late-type galaxies. The 
stellar content off of the M82 disk plane is also investigated. A mixture of bright 
main sequence stars, RSGs, and AGB stars are detected out to minor axis distances of 7 kpc. 
These stars, which span a range of ages, are concentrated along the outflow. 
The brightest extraplanar AGB stars define a system with an exponential scale height 
of $1.8 \pm 0.2$ kpc, as measured along the minor axis. It is suggested that the 
young and intermediate aged stars in the extraplanar regions formed in structures 
similar to M82 South, and that these were subsequently disrupted by the tidal action of M82.

\end{abstract}


\section{INTRODUCTION}

	The properties of most nearby large galaxies were almost certainly defined 
by mergers and interactions that occured during early epochs. 
Depending on factors such as the gas fraction of the progenitors, 
their mass ratios, and their relative orbital characteristics, 
mergers and tidal interactions can dramatically alter the morphologies of the 
initial systems (e.g. Barnes 1992; Mayer et al. 2001; Springel \& Hernquist 2005). 
Even in cases where the basic structural properties of the progenitors are not 
altered, galaxy-galaxy interactions can trigger enhanced levels of star formation 
(e.g. Larson \& Tinsley 1978, Iono, Yun, \& Mihos 2004, Kewley, Geller, \& Barton 2006), 
thereby altering the appearance of galaxies.

	Simulations predict that the bulk of the merger activity associated with the 
assembly of the Galaxy concluded $\sim 9$ Gyr in the past (Bullock \& Johnston 2005), and 
signatures of this eventful period may still be seen in the Galactic disk (e.g. 
Kazantzidis et al. 2007). Still, mergers and interactions continue to occur in the local 
universe, albeit at a more subdued level than during earlier epochs. 
As the nearest ensemble of galaxies in which there 
is evidence of on-going galaxy-galaxy interactions, the M81 group 
is an unprecedented laboratory for probing the impact of these encounters. 
Arguably the most dramatic evidence of interactions between M81 and its companions that 
has been presented to date comes from studies of neutral hydrogen. The HI around 
M81 and M82 has a disturbed morphology, and HI filaments link M81 with other members 
of the group (e.g. Brouillet et al. 1991; Yun, Ho, \& Lo 1994; Boyce et al. 2001). 
The distribution of gas suggests that a dramatic event occured, and timing arguements, 
in which the galaxy Ho IX is assumed to be a tidal fragment, suggest that M82 passed 
through the western disk of M81 $\sim 2 \times 10^8$ years in the past (Yun et al. 1994).

	The stellar content of M82 constitutes a fossil record that can be mined to probe 
the nature of the galaxy before it interacted with M81. Studies of the stellar content 
can also be used to measure the timing of the interactions with M81 by searching for past 
changes in the SFR. The vast majority of previous studies have focused on the 
central few kpc of M82, which is a region of intense star-forming activity. 
Dust largely obscures this part of M82, and so much work 
has focused on studies of star clusters in areas that have relatively low 
extinction (O'Connell et al. 1995), or of the integrated light. The star formation history 
deduced from star clusters (e.g. de Grijs, O'Connell, \& Gallagher 2001) and 
the visible integrated light spectrum (Mayya et al. 2006) 
suggest that elevated levels of star formation in M82 commenced some 0.5 
-- 1.0 Gyr ago. These studies also suggest that the spatial location of star-forming 
activity has changed with time, in the sense that the area 
of active star formation has contracted since the initial interaction with M81. 
Star formation at radii larger than $\sim 1$ kpc in the M82 
disk may have stopped some 0.5 Gyr in the past (Mayya et al. 2006), while the most 
recent star-forming activity in the inner regions of the disk occured only a few Myr ago 
(Gallagher \& Smith 1999; Forster Schreiber et al. 2003; Smith et al. 2006). 

	Given the compelling nature of the central star-forming regions 
it is perhaps not surprising that there have only been a few studies of 
the rest of M82. In what has been the largest areal study to date of resolved stars in M82, 
Sakai \& Madore (1999) used images recorded with WFPC2 to study red giant 
branch (RGB) stars in two adjoining fields in the eastern half of the galaxy. They 
find a distance modulus of 27.95 based on the brightness of the RGB-tip. A number 
of luminous candidate asymptotic giant branch (AGB) stars were also identified, hinting at 
star-forming activity during intermediate epochs.

	Davidge et al. (2004) used $H$ and $K'$ images 
recorded with NIRI $+$ ALTAIR on Gemini North to investigate the stellar content at 
a projected distance of $\sim 1$ kpc south of the M82 disk plane. 
The relative numbers of AGB and RGB stars were found to be 
consistent with stellar evolution models, while the peak brightness of the AGB is 
consistent with that of an intermediate age population. These data have an angular 
resolution of 0.08 arcsec FWHM, and the photometric properties of these highly evolved 
stars are not affected by crowding. These observations suggest that the region off of the 
M82 disk plane and in the outer disk was an area of recent star formation 
during the past $\sim 1$ Gyr, and contains a diverse stellar population.

	The extraplanar environment is another region of interest 
for stellar content studies. The starburst in M82 has triggered 
an outflow in which at least $3 \times 10^8$ M$_{\odot}$ 
of gas has been ejected from the galaxy (Walter, Weiss, \& Scoville 
2002). The material in the outflow may interact with clouds of gas and dust that surround 
M82, which presumably were ejected by earlier outflow events or were pulled from the 
galaxy during the interaction with M81. Devine \& Bally (1999) discuss the `Cap', 
which is located 11 kpc to the north of M82. The emission from the Cap is thought 
to be powered by the collision between the outflow and a gas cloud 
(Lehnert et al. 1999; Strickland et al. 2004). The Cap 
contains knots at visible wavelengths that may herald the early stages of star 
formation.

	Davidge (2008) investigated the stellar content of a stream of stars 
$\sim 6$ kpc south of M82. This feature, which was named `M82 South' by Sun et al. (2005), 
covers a $0.3 \times 3.1$ kpc$^2$ area and has an integrated brightness M$_V \sim -9.5$. 
Individual stars were resolved, and comparisons with 
isochrones indicate that star formation in M82 South last occured $\sim 50$ Myr ago. 
This age overlaps with the most recent era of large-scale star formation in tidal features 
throughout the M81 group (e.g. Makarova et al. 2002; Sakai \& Madore 2001). 
The oblong appearance of M82 South and the spatial flaring of stars at its eastern 
end suggest that it is dispersing, thereby raising the possibility that 
the young stars in M82 South may diffuse into the extraplanar regions of the galaxy.

	There has not yet been a comprehensive survey of the stellar 
content of the outer disk and extraplanar regions of M82. 
Such a survey should encompass an area covering at least a few hundred arcmin$^2$ 
since (1) the disks of late-type galaxies typically extend out to distances of at 
least 10 kpc, which translates to an angular distance of 9 arcmin at the 
distance of M82, and (2) the interface between the outflow and surrounding clouds 
occurs $\sim 10$ arcmin from M82 if the Cap is assumed to be a representative 
example of this phenomenon. In the current paper, the 
wide-field capabilities of the CFHT MegaCam and WIRCam imagers 
are used to conduct an initial reconnaisance of the stellar content 
in the outer disk and the extraplanar regions of M82.

	A combined visible $+$ infrared dataset 
offers significant benefits for investigating the stellar content of a galaxy like M82. 
The stellar types that dominate at visible and infrared 
wavelengths are very different. Not only does this enable a 
more comprehensive census of the most evolved stars, but it also provides a means 
of assessing the impact of crowding, as the contrast between the brightest objects and 
the underlying populations differ with wavelength. The impact of dust can also be 
investigated by comparing the photometric properties of stars that are sampled over a wide 
wavelength range. 

	A goal of this paper is to investigate the spatial extent and stellar 
content of the outer disk of M82, as this information provides insight into 
the nature of the galaxy before -- and immediately after -- its interaction with M81. 
The present data sample the semi-major axis of M82 out to the limiting
distances at which disk stars have been detected in non-interacting late type galaxies. 
These are regions where the disk surface brightness is too low to be detected in integrated 
light, but where individual disk stars can still be identified. The spatial extent of the 
M82 disk is of interest as Sofue et al. (1992) and Sofue (1998) discuss the 
rotation curve of M82, and suggest that the dark matter halo and outer disk of the galaxy 
were stripped away. This interpretation is not ironclad, and needs to be tested by directly 
measuring the size of the stellar disk. Indeed, a caveat when interpreting velocity 
measurements in M82 is that the dynamics in the central regions of the galaxy are 
dominated by a bar (Wills et al. 2000), which accounts for a significant fraction of the 
total galaxy mass, and so is a dynamically significant component (e.g. Greve et al. 2002). 
Moreover, the velocity field defined by the [SIII] 9069 \AA\ line challenges the 
presence of Keplerian rotation outside of the region affected by the bar 
(McKeith et al. 1993).

	As for the recent star-forming history of the 
outer disk of M82, if large-scale star formation outside of the 
central $\sim 1$ kpc of M82 terminated during intermediate epochs, as predicted by 
Mayya et al. (2006), then the CMDs of the outer disk should contain a 
prominent AGB and few -- if any -- supergiants or bright main sequence stars. 
The fractional contribution made by supergiants and main sequence stars 
to the total light in the outer disk of M82 should thus be smaller than in galaxies 
with on-going star formation. Moreover, if the star burst in M82 was originally global 
in nature, encompassing much of the disk before retreating 
to smaller radii, then the number density of AGB stars 
should also be higher at intermediate and large radii than in a normal late-type galaxy.
Is the stellar content of the outer disk of M82 consistent with these expectations?

	Another goal of this paper is to investigate the nature and distribution 
of the brightest objects in the extraplanar regions of M82, as these 
also provides clues into M82's past and future. If M82 was 
a late-type galaxy prior to interacting with M81 then it would have had a modest 
entourage of globular clusters. These clusters are the brightest tracers 
of the early history of M82, and the discovery of even a modest globular cluster population 
in the extraplanar regions would suggest that the interaction with M81 did not strip 
the outermost stellar regions from the galaxy. In addition, if other stellar collections 
like M82 South have formed in the recent past then a significant stellar component 
made up of young and/or intermediate age stars may be present off of the disk plane. The 
spatial distribution of such stars provides clues into their places of origin. Moreover, if 
the extraplanar regions are populated with stars that span a range of ages and 
metallicities then an observer a few Gyr in the future will see M82 as a galaxy with a 
diverse halo population, as is seen in the halos of some nearby late-type galaxies.

	The paper is structured as follows. The acquisition and processing 
procedures are summarized in \S 2, along with a description of the photometric 
measurements. The photometric properties and spatial distribution of stars in the outer disk 
and extraplanar regions of M82 are discussed in \S\S 3 and 4. 
A summary and discussion of the results follows in \S 5.

\section{OBSERVATIONS AND REDUCTIONS} 

\subsection{MegaCam}

	Deep images of a single MegaCam (Boulade et al. 2003) 
pointing centered midway between M81 and M82 were 
recorded through $r', i'$ and $z'$ filters on the 
night of October 23 UT 2006. The detector in Megacam is a mosaic of 
thirty six $2048 \times 4612$ pixel$^2$ CCDs, that together cover 
an area of one degree$^2$ with 0.185 arcsec pixel$^{-1}$. 
The data were recorded with a square-shaped dither pattern to assist with 
the identification of bad pixels and the suppression of 
cosmic rays. Four 300 sec exposures were recorded in $r'$ and $i'$, 
while four 500 sec exposures were recorded in $z'$. Stars have 0.7 -- 0.8 arcsec 
FWHM in the final processed images, depending on the filter.

	The removal of instrumental and environmental signatures from the raw images
was done with the ELIXIR package at the CFHT. This processing included bias subtraction, 
flat-fielding, and the subtraction of a fringe frame. The ELIXIR-processed images 
were then aligned, stacked, and trimmed to the area that is common to all exposures.

\subsection{WIRCam}

	Images of M82 were recorded through $J, H$ and $Ks$ filters 
with WIRCam (Puget et al. 2004) on the night of February 4 UT 2007. The detector in 
WIRCam is a mosaic of four $2048 \times 2048$ HgCdTe arrays. 
Each exposure covers $21 \times 21$ arcmin$^2$, with 0.3 arcsec pixel$^{-1}$.

	The WIRCam data were recorded with a 
square-shaped dither pattern. A total of 20 45 sec exposures were recorded in 
$J$, while 120 15 sec exposures were recorded in $H$, and 
160 15 sec exposures were recorded in $Ks$. The $J$ data were obtained 
over one dither cycle (i.e. five 45 sec exposures were recorded per dither position), 
while the $H$ and $Ks$ data were obtained over two dither cycles. 
Stars in the final images have 0.8 arcsec FWHM.

	The initial processing of the WIRCam data 
was done with the I`IWI pipeline at the CFHT, and this 
consisted of dark subtraction and flat-fielding. To continue the processing, a 
calibration frame to remove interference fringes and thermal emission artifacts 
was constructed by median-combining the I'IWI-processed M82 images with those of an 
adjacent field that were recorded immediately after M82. The data were DC sky-subtracted 
before they were combined to account for exposure-to-exposure variations in sky brightness. 
In addition, a pixel-by-pixel low-pass clipping algorithm was applied 
to suppress stars and galaxies in both fields. The resulting calibration frames were 
subtracted from the flat-fielded data. The processed images were then aligned,
stacked, and trimmed to the area that is common to all exposures.

\subsection{Photometric Measurements}

	The photometric measurements were made with the point-spread-function 
(PSF) fitting program ALLSTAR (Stetson \& Harris 1988). The PSF and 
star catalogues used by ALLSTAR were obtained from DAOPHOT (Stetson 
1987) routines. The photometric catalogues were culled in two ways to reject objects 
with poorly determined photometric measurements. First, all objects in which the fit 
error ($\epsilon$) computed by ALLSTAR exceeded 0.3 mag were rejected. This 
removed objects near the faint limit of the data, where photometry is problematic. Second, 
objects that have an $\epsilon$ that is markedly higher than the majority of objects 
with the same brightness were also removed. The objects rejected 
in this step tend to be either (1) non-stellar, (2) multi-pixel cosmetic defects, 
and/or (3) in the crowded inner regions of M82.

	The photometric measurements were calibrated using 
standard star observations that were obtained within a few days of the science 
observations. The MegaCam photometry were calibrated using the zeropoints that are 
placed in the data headers during ELIXIR processing, which are computed 
from standard star observations that are obtained during each 
MegaCam observing run. The WIRCam observations were calibrated using zeropoints 
published on the CFHT website that were computed from standard star observations 
made during February 2007.

	Artificial star experiments were used to estimate sample completeness,
the random scatter in the photometry, and the brightness at which blends 
of fainter stars account for a significant number of objects. The artificial stars were 
assigned brightnesses and colors that are consistent with the main locus of stars in 
M82. As with actual sources, an artificial star was considered to be recovered only if 
it was detected in at least two filters. The artificial star measurements were subjected 
to the $\epsilon$-based rejection criterion discussed previously.

	The results of the artificial star experiments depend on the stellar density, in 
the sense that as stellar density drops then at a given magnitude (1) the completeness 
fraction increases, (2) the random scatter in the photometry decreases, 
and (3) the incidence of interloping blends of fainter stars decreases. For 
sources in the M82 disk plane, the 50\% completeness level 
in the MegaCam measurements occurs near $i' = 24.0$ 
in the 4 -- 6 kpc distance interval and $i' = 24.5$ in the 10 -- 12 kpc interval. 
As for the WIRCam measurements, the 50\% completeness level occurs near $K = 19.8$ 
in the 4 -- 6 kpc distance interval, and $K = 20$ in the 10 -- 12 kpc interval. The 
completeness fraction is less sensitive to changes in stellar density in the WIRCam data 
than in the MegaCam data because of the dimished impact of line blanketing at wavelengths 
longward of $1\mu$m, which results in greater contrast between AGB stars and the 
main body of red giant branch and bluer stars in the infrared than the visible. 
The completeness fractions of stars in the extraplanar regions 
are comparable to those in the 10 -- 12 kpc disk interval. The artificial 
star experiments indicate that blends can constitute a significant fraction of stars 
at magnitudes where the completeness fraction is below $\sim 50\%$.

\section{THE OUTER REGIONS OF THE M82 STELLAR DISK}

	Based on its total mass, luminosity, and stellar content, O'Connell \& 
Mangano (1978) argue that M82 was a late-type disk or irregular galaxy prior to 
interacting with M81. Morphologically, the spiral structure of M82 
is consistent with type SBc (Mayya, Carrasco, \& Luna 2005), although 
there may be structural differences with respect to late-type galaxies in the 
local Universe. Indeed, if gas in the disk of M82 defines a Keplerian rotation curve 
out to a radius of $\sim 4$ kpc (e.g. Sofue et al. 1992; Sofue 1998 -- 
but also the discussion in \S 1) then M82 lacks (1) a massive halo 
and (2) an extended stellar disk.

	While the observational properties of M82 have undoubtedly been affected by 
interactions with M81, signatures of its nature prior to this encounter may remain. 
In this section we seek (1) to measure the spatial extent of the 
M82 stellar disk, and determine if it is comparable to that in 
other nearby late-type spirals, which have been traced out to many scale lengths 
(e.g. Davidge 2007; Bland-Hawthorn et al. 2005), and (2) 
to determine if the spatial density of old and intermediate age 
stars is comparable with that of other late-type galaxies.
Only stars that are within $\pm 65$ arcsec of the major axis of M82 
are considered. While this criterion excludes stars along the minor axis 
of M82 that belong to the disk, it reduces contamination from extraplanar stars, some of 
which appear to have young and intermediate ages (\S 4). 

\subsection{An Overview of the CMDs}

	The $(i', r'-i')$ and $(z', i'-z')$ CMDs of stars within $\pm 65$ arcsec 
of the major axis of M82 are shown in Figure 1. The stars have been grouped 
according to projected galactocentric distance in the disk plane, assuming a distance 
modulus of 27.95 (Sakai \& Madore 1999) and a disk inclination of 77 degrees (Mayya 
et al. 2005). A consequence of the large disk inclination is that the various 
radial intervals sample comparable projected areas on the sky, and so the number of 
contaminating foreground stars and background galaxies is roughly the same in each interval.

	The plume of objects with $r'-i'$ between 0 and 1 and $i' < 22$ in the top 
row of Figure 1, and with $i'-z' \sim 0$ and $z' < 22$ in the bottom row,
consists of Galactic disk stars, although some of the 
blue objects with $i' \sim z' > 20$ in the 2 -- 4 kpc CMDs could be 
young and intermediate age star clusters in M82. 
Stars in M82 dominate the lower half of the CMDs in Figure 1. 
The concentration of objects that peaks near $i' \sim 22.5 - 23$ and $z' \sim 
22 - 22.5$ in the R$_{GC} > 4$ kpc CMDs is made up of stars that are evolving on the AGB, 
and so have ages $> 100$ Myr. The AGB sequence is more vertical on the $(z', i'-z')$ 
CMDs than the $(i', r'-i')$ CMDs because of the diminished impact of line blanketing at 
longer wavelengths. The concentration of AGB stars can be seen in the MegaCam CMDs out 
to at least R$_{GC} \sim 10$ kpc, hinting that the stellar disk extends 
to at least this radius. Background galaxies, which span a range of colors and 
increase in numbers towards fainter magnitudes, are the largest source of 
contamination at magnitudes and colors of the AGB component, and make the 
identification of individual AGB stars in the CMDs at larger radii difficult. 
The issue of the radial extent of the M82 disk is examined in 
greater detail in \S 3.4, where it is demonstrated that a statistically significant 
excess number of disk AGB stars can be traced out to R$_{GC} = 12$ kpc.

	The $(K, H-K)$ and $(K, J-K)$ CMDs of stars along 
the major axis of M82 are shown in Figure 2. As with the CMDs at shorter wavelengths, 
foreground Galactic disk stars form a plume of relatively blue 
objects in the top half of the CMDs. Stars in M82 occur in large 
numbers when $K \geq 19$ and -- as with the CMDs in Figure 1 
-- the majority of these are evolving on the AGB. 
The peak brightness of the AGB in these CMDs ($K \sim 18.5 - 19$) 
is well above the magnitude where contamination from blends of fainter stars is 
expected to be an issue (\S 2.3). As in Figure 1, a concentration of AGB stars 
can be seen in the near-infrared CMDs out to at least R$_{GC} = 10$ kpc.

	The ridgeline of the bright AGB sequence in M82 has $H-K \sim 0.4$ and 
$J-K \sim 1.3$. Normal background galaxies at cosmologically modest redshifts
tend to have redder colors than these because the 
first overtone CO bands, which are among the strongest features in the spectrum of 
highly evolved stars and have a large influence on $J-K$ and $H-K$ 
colors, are shifted out of the $K$ bandpass when $z \geq 0.1$. 
The mean redshift of galaxies at $K \sim 17$ is $<z> \sim 0.4$ (Cowie et al. 
1996), and a normal galaxy at this redshift will have $H-K \sim 0.8$ and 
$J-K \sim 1.6$ after applying the k-corrections from Mannucci et al. (2001). 
In fact, collections of objects with such colors are 
seen in the CMDs of the two outermost intervals in Figure 2. 
There are objects with $H-K > 0.5$ and $J-K > 1.5$ in the CMDs that sample smaller 
galactocentric distances in Figure 2, and some of these may be C stars in M82 (e.g. 
Davidge 2005). However, the reader is cautioned that some of these 
objects are probably background galaxies rather than stars in M82.

\subsection{Comparisons with Isochrones}

	Isochrones with ages log($t_{yr}$) = 7.5, 8.0, 8.5, and 9.0 
from Girardi et al. (2002; 2004) are compared with selected 
$(i', r'-i')$ and $(K, J-K)$ CMDs in Figures 3 
(Z=0.008) and 4 (Z=0.019). A distance modulus of 27.95 (Sakai 
\& Madore 1999) is assumed, with A$_B = 0.12$ mag (Burstein \& Heiles 
1982). No correction for internal reddening has been applied.

	The reddening maps of Schlegel, Finkbeiner, \& Davis (1998) indicate that 
A$_B = 0.69$ mag for M82, which is considerably higher than the 
foreground extinction measured by Burstein \& Heiles (1982). This difference is undoubtedly 
due to emission from dust in M82. The dust in M82 is centrally concentrated, and 
is not expected to be a major influence at large radii. Still, the comparisons in 
Figures 3 and 4 are not greatly affected by internal extinction as long as the 
true distance modulus also accounts for any additional 
source of reddening. For example, if A$_B = 0.69$ is used to 
compute a revised `true' distance modulus from the Sakai \& Madore (1999) RGB-tip 
measurement, which is based on observations of stars in the outer disk, 
then the ages inferred from comparisons with the isochrones 
differ by only $\sim 0.1$ dex from those measured if A$_B = 0.12$ mag is assumed.

	The ages inferred for the main locus of 
AGB stars from the Z = 0.008 and Z = 0.019 sequences differ by 0.5 dex. 
While the interstellar medium (ISM) and young stars in M82 have metallicities approaching 
(e.g. Martin 1997; Gallagher \& Smith 1999) and even exceeding (Origlia et al. 2004) solar, 
the metallicities of stars that formed during intermediate epochs are 
almost certainly lower than those of younger stars. Parmentier et al. (2003) 
estimated metallicities for clusters in M82, and 
these data indicate that clusters with ages log(t) = 8.0 have slightly sub-solar 
metallicities, while clusters with log(t) = 8.5 have Z = 0.008 
(e.g. Table 1 of Davidge et al. 2004). The Z = 0.008 models are 
adopted as the baseline for subsequent discussion and comparison, although 
the reader is reminded that the AGB age estimates depend on the adopted metallicity.

	The Z = 0.008 isochrones predict that the majority of stars in the $(i', 
r'-i')$ CMD have ages between log(t) = 8.0 and 9.0. However, there is also a modest 
population of objects with $(r'-i') \sim -0.3$ that have photometric properties that are 
consistent with them being main sequence stars with ages log(t) $\sim 7.5 - 8.0$. 
These stars have brightnesses where contamination from blends of fainter stars 
becomes significant (\S 2.3), and so the detection of moderately bright main 
sequence stars in these data should be considered to be preliminary. Still, 
evidence to support the main sequence interpretation comes from the presence 
of possible red supergiants (RSGs) that are located on or near isochrones that 
pass through the region of the CMDs that is occupied by the candidate main sequence stars.

	The benefits of observing metal-rich AGB stars at wavelengths 
longward of $1\mu$m are clearly illustrated when comparing the isochrones 
in Figures 3 and 4. The models show that the AGB slumps over on the 
$(i', r'-i')$ CMD, whereas the AGB is much closer to vertical on the $(K, 
J-K)$ CMD, due to the diminished impact of line blanketing at longer wavelengths. 
In fact, the locus of peak M$_K$ brightnesses defined by the log(t) = 8.0, 8.5, and 9.0 
isochrones roughly tracks the upper envelope of the AGB concentration 
on the $(M_K, J-K)$ CMD. This being said, the isochrones only pass through the blue end of 
the AGB plume, and miss many red sources in M82. This situation does not change when the 
Z = 0.019 models are considered. The inability to reproduce the full range of observed 
near-infrared colors may be due in part to incomplete -- as opposed to incorrect -- physics 
in the models. At least some of the objects with $J-K \geq 1.5$ may be C stars 
(e.g. Davidge 2005), and the Girardi et al. (2002; 2004) models do not include these objects.
It should also be kept in mind that this is the color range where the number density 
of background galaxies is highest (\S 3.1), so at least some of the
red objects probably do not belong to M82.

\subsection{An Overview of the LFs}

	The $i'$ LFs of objects with $r'-i'$ between 0 and 2 
are shown in Figure 5. The LFs have been corrected for contamination 
from foreground stars and background galaxies by subtracting the LF of objects 
in control fields to the east and west of M82. The control fields subtend much larger 
areas than the M82 disk fields, and the control field LFs were scaled to account 
for these differences in area. The effects of crowding are clearly 
evident at the faint end of the 2 -- 4 and 4 -- 6 kpc LFs, where incompleteness 
causes the inflexion point in the number counts to set in 
$\sim 0.5$ magnitude brighter than at larger radii. This is consistent with the 
results of the artificial star experiments (\S 2.3).

	After correcting statistically for foreground and background objects, 
an excess number of stars remains in the $10 - 12$ kpc interval, 
indicating that the stellar disk of M82 extents out to at least this radius. Therefore, 
while the CMDs that cover the 10 -- 12 kpc intervals in Figures 1 and 2 do 
not show an obvious concentration of AGB stars, a modest number of stars that belong to M82 
are present. The peak AGB $i'$ magnitudes predicted for various ages from 
the Z = 0.008 Girardi et al. (2004) models are indicated at the top of 
Figure 5. Caution should be exercised when using this calibration, as photometric 
variability smears the location of the AGB-tip. There are also 
uncertainties in the models, which only extend to the onset of thermal pulses. 
These models also overestimate the number of AGB stars when compared with RGB 
stars, possibly indicating that mass loss rates have been underestimated 
(Williams et al. 2007). With these caveats in mind, there is evidence of 
a log(t) $= 8.1$ population in all distance intervals, indicating 
that much of the M82 disk formed stars during intermediate epochs. 
The objects with $i' < 22$ in the $i'$ LFs are RSGs that formed within the 
past $\sim 0.1$ Gyr.

	The $K$ LFs of objects with H--K between 0 and 1 are shown in Figure 6. 
As with the $i'$ LFs, the $K$ LFs indicate that 
the stellar disk of M82 extends out to at least 12 kpc, and that intermediate age stars are 
present over a large fraction of the stellar disk. The number of stars per magnitude 
in the $K$ and $i'$ data differs because the AGB is distributed over 
a larger range of $K$ magnitudes than $i'$ magnitudes. The AGB-tip at 10 -- 12 kpc 
in the $K$ LF is consistent with an age log(t) = 9.0, which is 
older than the AGB-tip brightnesses at smaller radii. While this points to a possible
conflict with the  $i'$ LFs, where the AGB-tip brightness corresponds to log(t) = 8.1 
at all radii, there are large uncertainties in the number counts at the bright end in the 
10 -- 12 kpc LFs in both filters, so the significance of any difference is low. 
The population of RSGs that is seen in the 2 -- 6 kpc $i'$ LFs appears to be missing 
in the $K$ LFs. However, this is likely a consequence of the relative 
brightnesses of RSGs and AGB stars in the near-infrared, which is such that 
RSGs with ages $> 60$ Myr are fainter in $K$ than 
AGB-tip stars that have log(t) $< 9.0$.

\subsection{The Spatial Distribution of Stars in the Disk}

	Specific frequency (SF) measurements provide a means of quantifying the spatial 
distribution of stars. Following previous studies by the author, the SF is taken to be the 
number of objects in a given color interval that would be seen in an object with an 
integrated brightness M$_K = -16$. The total M$_K$ in each spatial interval in M82 was 
calculated from the $J-$band light profile measured by Jarrett et al. (2003) from 2MASS 
data, which was extrapolated along an exponential profile to include the outermost 
regions of the disk. While $K-$band measurements are more sensitive 
to the stars that trace most of the stellar mass, the 2MASS $K-$band light 
profile does not extend to radii larger than $\sim 8$ kpc in M82, and at large radii is much 
noisier than the $J-$band profile. M$_K$ in each interval was 
calculated by assuming that J--K = 0.8, in agreement with 2MASS observations of M82 at 
intermediate radii. 

	The SFs of red stars in the MegaCam and WIRCam data are shown in 
Figures 7 and 8. The error bars show the uncertainties due to the number of objects in each 
magnitude interval. The dotted line is a power law that was fit to the mean SF measurements 
at intermediate brightnesses in the 4 -- 10 kpc intervals, and this relation is shown 
in the figures to provide a reference for annulus-to-annulus comparisons. The $i'$ and $K$ 
SF measurements agree with the mean relation throughout much of the disk. 
The one exception is in the 2 -- 4 kpc interval, where SF$_{i'}$ is 
lower than the mean relation when $i' > 22$. This is due to incompleteness in 
the MegaCam data in this crowded portion of the galaxy, 
and the SF$_K$ measurements at the same radii agree with the mean trend. 
The comparisons in Figures 7 and 8 indicate that the AGB component throughout 
most of the disk of M82 is uniformly mixed with the main body of stars that dominate 
the integrated light.

	Comparing the SFs of stars in M82 
with those in other galaxies is of interest for 
probing the nature of M82 prior to the interactions with M81. 
The star-forming history of M82 can be explored in a purely empirical 
manner by comparing the SF measurements in Figures 7 and 8 with those in 
other galaxies. If the specific SFR in the outer disk of M82 
was greatly elevated during intermediate epochs then the SF of AGB stars 
in M82 should be higher than in a `normal' star-forming galaxy. 
The Sc galaxy NGC 2403 is adopted here as a benchmark `normal' late-type spiral galaxy 
for this comparison, as it is in an isolated part of the M81 group (e.g. Karachentsev 
et al. 2002), has a stellar content that is similar to that of 
other late-type spirals (e.g. Davidge \& Courteau 2002), 
and has moderately deep MegaCam and WIRCam observations (Davidge 2007). 

	The SF of stars in NGC 2403 were computed from the observations discussed by 
Davidge (2007) after applying the same selection criteria employed to reject 
objects with large photometric uncertainties in M82 (\S 2.3). 
The light profile of NGC 2403 measured from 2MASS data does 
not extend to large radii, and so the $K-$band surface 
brightnesses in NGC 2403 were computed from the $r-$band surface brightness profile that 
was `extended by hand' by Kent (1987). A color $V - K = 2.2$ was assumed, 
based on measurements of NGC 2403 at radii where the 2MASS 
and $r-$band light profiles overlap. 

	The SFs of stars in the outer disk of M82 and NGC 2403 are compared in Figure 
9. The NGC 2403 SF measurements have been shifted faintward by 0.5 magnitudes as the distance 
modulus of NGC 2403 (Freedman \& Madore 1988) is 0.5 mag lower than that of M82.
As in Figures 7 and 8, the error bars show the uncertainties due to the number of 
objects in each magnitude interval. After correcting for differences 
in distance, the $i'$ SFs of red stars in M82 and NGC 2403 
are in reasonable agreement at $i' = 23.0 - 23.5$, which is where AGB-tip stars with ages 
$\sim 0.3 - 0.6$ Gyr are found. As for the $K-$band data, 
the SF measurements of both galaxies agree at K= 19, which corresponds to the peak brightness 
of AGB stars with ages $\sim 0.6$ Gyr. However, the NGC 2403 SF$_K$ curve falls 
below the M82 measurements at fainter magnitudes, suggesting that 
the specific SFRs of the outer disks of M82 and NGC 2403 may have differed 
$> 1$ Gyr in the past, in the sense that M82 had more vigorous star-forming activity.

	The situation is different when the SF measurements of NGC 2403 and M82 are 
compared at the bright end. The $K-$band SF measurements of NGC 2403 fall above those 
of M82 when K$_{M82} < 19$. While the large error bars indicate 
that caution should be exercised when drawing 
conclusions from the SF$_{i'}$ comparisons alone, when $i'_{M82} < 23$ 
there is a systematic tendency for the NGC 2403 SF$_{i'}$ measurements to fall above those 
of M82. These differences occur in the magnitude range that contain RSGs and the brightest 
AGB stars (Figure 3). Therefore, the comparisons in Figure 9 
suggest that while star formation has continued in the outer disk of NGC 
2403 during the past $\sim 100$ Myr, the specific SFR has been lower in the 
outer disk of M82 during this same time period. 

	If the specific SFRs in the outer disks of NGC 2403 and 
M82 during the past few hundred Myr have been markedly 
different then this should be evident when their CMDs are compared. 
The $(i', r'-i')$ CMDs of stars with galactocentric distances 
between 6 and 8 kpc in the disks of NGC 2403 and M82 are compared in Figure 10. 
As in Figure 9, the NGC 2403 measurements have been shifted faintward by 0.5 magnitude 
to adjust for the difference in distance. There are many more stars with R$_{GC}$ between 
6 and 8 kpc in NGC 2403 than in M82, and this complicates efforts to compare the CMDs. 
To account for this, the middle panel of Figure 10 shows the CMD of stars 
in the NGC 2403 6 - 8 kpc interval in small fields near the minor and 
major axes of NGC 2403 that together contain the same number of stars with $i'_{M82}$ 
between 23.25 and 23.75 as in the 6 -- 8 kpc interval in M82.

	The CMDs of NGC 2403 and M82 in Figure 10 are very different. The M82 CMD 
lacks the plume of RSGs with $r'-i'$ between 0.3 and 1.1 
that dominates the NGC 2403 CMD when $i'_{M82} < 23$ (i.e. 
M$_{i'} < -5$). The isochrones in Figure 3 indicate that the RSG 
sequence in this magnitude range is dominated by stars that are younger than 
$\sim 100$ Myr. The CMD of M82 also lacks the spray of bright blue main 
sequence stars with $r'-i' < 0$ that is seen in the NGC 2403 CMD. 
This is also consistent with the outer disk of M82 lacking stars that 
formed during the past $\sim 100$ Myr. In summary, while the 
comparisons in Figure 9 indicate that the specific SFRs in 
the outer disks of M82 and NGC 2403 may have been similar 0.1 -- 1 Gyr in the past, 
the CMDs in Figure 10 indicate that during the past $\sim 100$ Myr the specific SFR in the 
outer disk of M82 has been much lower than in NGC 2403. 

\section{BRIGHT EXTRAPLANAR STARS AND CLUSTERS}

	The outflow from the star-forming regions in M82 injects hot 
gas into the extraplanar environment. Some of this 
gas appears to have cooled sufficently to allow stars to form, as stars 
with ages $\sim 50$ Myr are seen in M82 South (Davidge 2008). While M82 
South appears to be the most obvious extraplanar stellar structure associated with M82 
(\S 4.2), there may be other, albeit less pronounced, collections of young extraplanar 
stars. In this section, a search is conducted for young and intermediate age stars in the 
extraplanar environment. A list of globular cluster candidates is also compiled.

\subsection{An Overview of the CMDs}

	The investigation is restricted to a 6.5 arcmin wide strip that extends perpendicular 
to the major axis of M82. Given that the stellar disk of M82 has been traced out to a 
galactocentric radius of {\it at least} 12 kpc (\S 3), and that 
the disk of M82 is inclined at 77 degrees, then disk stars will be present out to 
a projected distance along the minor axis of at least $\sim 3$ kpc. 
Therefore, to reduce contamination from disk stars on the near and far sides of the 
M82 disk only projected minor axis distances, D$_Z$, in excess of 2.5 kpc are considered. 
An ellipticity of 0.68 (Jarrett et al. 2003) is assumed when sorting stars into D$_Z$ 
intervals.

	The $(i', r'-i')$ and $(z', r'-i')$ CMDs of the extraplanar 
regions of M82 are shown in Figure 11, while the corresponding $(K, H-K)$ and $(K, 
J-K)$ CMDs are shown in Figure 12. The CMDs that are closest to the disk plane are richly 
populated, with a conspicuous concentration of AGB stars. There is also a spray 
of bright red objects above the AGB, many of which are probably RSGs. 
Contamination from disk stars is greatest at D$_Z = 
3$ kpc, and so it is perhaps not surprising that the D$_Z = 3$ kpc CMDs are similar to 
those in the outer disk of M82 (\S 3.1). 

	The number of stars that belong to M82 decreases as D$_Z$ increases, and 
at D$_Z = 7$ kpc contamination from relatively blue ($H-K \sim 0.1$, 
$J-K \sim 0.6$) foreground stars and relatively red ($H-K \sim 0.9$, 
$J-K \sim 1.6$) background galaxies dominate the infrared CMDs in Figure 12. Still, a 
concentration of AGB stars can be traced out to D$_Z = 7$ kpc in Figures 11 and 12, while 
objects with colors and brightnesses that are consistent with them being 
RSGs are also seen out to D$_Z = 7$ kpc in Figure 11. 
The presence of possible RSGs at this D$_Z$ is perhaps not surprising given 
that RSGs are present in M82 South (Davidge 2008). 

\subsection{Comparisons With Isochrones}

	The CMDs of objects with D$_Z = 3$ and 5 kpc are compared with isochrones from 
Girardi et al. (2002; 2004) in Figure 13. Given that gas in 
M82 appears to have a solar metallicity (discussion in \S 3), then Z = 0.019 
models are shown in Figure 13. These models do a reasonable job of 
matching the red envelope of objects with M$_{i'} > -5$ in the MegaCam data, although 
some of the reddest objects are probably background galaxies. It should be recalled 
that the Z = 0.008 and Z = 0.019 isochrones predict ages for AGB stars that differ 
by $\sim 0.5$ dex (\S 3).

	There are objects in the $(M_{i'}, r'-i')$ CMDs that have photometric properties 
that are consistent with those of main sequence stars with ages $\leq 100$ 
Myr. These are likely real sources, as opposed to spurious blends of 
objects, for two reasons. First, the sequence of 
blue stars extends well above the $i' = 24.5$ (M$_{i'} = -3$) 
limit where the artificial star experiments suggest that 
blends occur in significant numbers (\S 2.3). Second, the region of the CMDs that contains 
RSGs of the same age as the candidate main sequence stars is well populated. 
Older objects are also present, as there are stars on the AGB that have an age 
log(t) $\leq 8.5$ on the (M$_{i'}, r'-i')$ CMDs. Thus, it is concluded that 
the extraplanar regions of M82 contains stars (1) with ages $< 100$ Myr, and (2) 
that span a range of ages.

\subsection{An Overview of the LFs}

	The $i'$ LFs of stars with $r'-i'$ between 0 and 2 are shown in Figure 14, 
while the $K$ LFs of stars with $H-K$ between 0 and 1 are shown in Figure 15. 
The LFs were corrected for contamination from foreground stars and background galaxies by 
subtracting number counts from the control fields, which were scaled to match the 
counts expected in the area covered at each D$_Z$. There remains a significant number 
of stars at the faint end of the LFs in all D$_Z$ intervals after applying this correction, 
quantitatively confirming the results from the 
CMDs in Figures 11 and 12 that intermediate age stars with ages $\leq 1$ Gyr 
are seen in the extraplanar regions of M82. 

	The dashed lines in Figures 14 and 15 show the reference SF relations from Figures 7 
and 8, which have been scaled to match the number of 
objects near the faint end of the extraplanar LFs. These fiducial relations are in 
reasonable agreement with the general trends defined by the extraplanar LFs. 
Still, there is an excess number of objects with $K = 18.5 - 19.0$ in the 
D$_Z = 7$ kpc interval when compared with the fiducial relation, and also 
near $i' = 21$ in the D$_Z = 6$ and 7 kpc intervals. 
These departures from the disk trend are due to RSGs in M82 South. 

\subsection{The Spatial Distribution of Extraplanar Sources}

	The numbers of objects in the $i' = 23$, 23.5, and 24.0 bins of Figure 
14, which are the magnitude intervals that are dominated by bright AGB stars, 
indicate that the exponential scale length of extraplanar AGB stars along the minor axis 
is $1.8 \pm 0.2$ kpc. This is comparable to the scale lengths of stellar 
disks. The spatial distribution of sources with $r'-i' 
< -0.2$ and $i' < 25$, which are notionally main sequence stars, and sources 
with $r'-i'$ between 0 and 2, and $i'$ between 24.5 and 23.0, 
which are candidate AGB stars and RSGs, are shown in Figure 16. 
It can be seen that the spatial distribution of sources is not equally distributed between the 
northern and southern portions of the galaxy. 

	There is a tendency for the blue objects to have higher densities close to the M82 
disk, and there are significantly more blue objects to the south of M82 than to the north. 
Blue objects are seen out to $\sim 7$ kpc, especially to the 
south of the main body of the galaxy. M82 South, which contains stars with an 
age $\sim 50$ Myr (Davidge 2008), appears as only a modest collection of 
blue objects in Figure 16.

	There are many more red sources than blue sources, making candidate AGB stars 
potentially more powerful probes of structure. The candidate AGB stars tend 
to occur along the minor axis in the outflow. M82 South is clearly visible in the 
distribution of red stars as the collection of objects near (x,y) = (2600,6000) 
in the co-ordinate system used in Figure 16. M82 South 
roughly defines the southern boundary of the concentration of AGB stars that are 
associated with the southern outflow. There are no other concentrations of AGB stars 
that are similar in density to M82 South; M82 South could be a unique object, or 
if similar structures formed earlier then they have since dispersed.

\subsection{Fossil Globular Clusters}

	If, as suggested by O'Connell \& Mangano (1978), M82 was at one time a late-type 
disk galaxy then it also would have been acompanied by a modest entourage of globular 
clusters. The number of clusters that would have been present can be estimated 
by assuming that the SF of globular clusters in M82
was similar to that of other late-type galaxies. Given the current levels of star 
formation in M82, the total brightness of M82 is best compared with that of other galaxies 
in the near-infrared, where the signal is less sensitive to differences in stellar content 
and dust extinction. With a total brightness $K = 4.7$ (Jarrett et al. 2003), then M$_K \sim 
-23.3$ for M82. The Leitherer et al. (1999) STARBURST99 models 
suggest that even if a large fraction of the $K-$band light comes from recently formed 
stars, which is likely the case near the center of M82, then the $K$ brightness 
will fade by no more than $\sim 2$ mag over the next billion years. Thus, it is likely 
that M$_K < -21.3$ for M82 in its pre-interaction state. The integrated brightness of the Sc 
galaxy NGC 2403 is M$_K \sim -21.3$ (Jarrett et al. 2003), and it is accompanied by tens of 
globular clusters (e.g. discussion in Davidge 2007). Therefore, the progenitor of M82 
then likely also had tens of old clusters if it was a late-type disk galaxy.

	While some of the globular clusters associated with M82 may have been lost during 
the encounter with M81, a remnant fossil population might remain. 
Saito et al. (2005) discuss a combined imaging and spectroscopic survey for 
globular clusters within 3 arcmin of the center of M82. They find 40 objects with brightnesses
and colors that fall within the range occupied by Galactic globular clusters, and a 
spectroscopic follow-up of a subset of these yielded 2 old globular clusters 
and 3 young clusters. 

	The ironclad identification of clusters outside 
of the area studied by Saito et al. (2005) will require 
spectroscopic information and high angular resolution images, but photometric data alone 
with sub-arcsec resolution can still be used to identify plausible globular cluster 
candidates. The WIRCam observations of M82 are of particular interest in this regard 
because they cover a large field of view, and metal-poor globular clusters have colors 
that are different from the vast majority of bright stars in galaxy disks. 
Following the procedure employed by Davidge (2007), candidate old globular clusters in 
the extraplanar regions of M82, as defined in \S 4.1, were identified based 
on their infrared colors, angular size, and appearance. More specifically, objects 
were identified that (1) have $J-K \leq 1$, and (2) are non-stellar, based on the 
DAOPHOT {\it sharp} parameter. The objects identified with these criteria were then 
inspected by eye to remove sources that are obvious spiral galaxies or blended stars. 
The crowded central regions of the galaxy, where the largest number of 
clusters may be located, was avoided as stellar blends and asterisms may 
masquerade as clusters in such an environment.

	If the globular cluster LF (GCLF) of the progenitor galaxy was like that in 
M31 (Barmby, Huchra, \& Brodie 2001) then the majority of globular clusters in M82 will have 
M$_K$ between --12 and --9, or $K \sim 16 - 19$, and should thus be bright enough to be 
detected in the WIRCam data. Five globular cluster candidates were found, and their locations, 
brightnesses, colors, and de-convolved angular sizes, where the latter assumes that the seeing 
disk and light distribution of the candidate clusters are Gaussians, are summarized 
in Table 1. The astrometry is based on the co-ordinates of stars detected 
by 2MASS close to the cluster candidates. All of 
the candidate clusters are to the south of M82, and three are within 5 arcmin of each 
other at a projected distance of $\sim 17$ kpc (15 arcmin) from the disk plane.

	The intrinsic angular sizes of the 
cluster candidates corresponds to $7 - 13$ parsecs at the distance of M82. 
This places these objects at the upper end of the half-light radius distribution of 
old Galactic halo globular clusters, but well within the range of half light radii of 
objects identified as young halo clusters by MacKey \& van den Bergh (2005). 
The criterion used by MacKey \& van den Bergh (2005) to distinguish 
between old and young halo clusters is horizontal branch 
morphology, and the young halo clusters are only a few Gyr younger 
than the oldest halo clusters. Thus, despite being labelled 
as `young' they still formed during relatively early epochs. 

\section{DISCUSSION AND SUMMARY}

	Wide field CFHT MegaCam and WIRCam images with sub-arcsec angular resolution 
have been used to probe (1) the spatial extent and stellar content of the M82 
stellar disk, and (2) the stellar content in the extraplanar regions of M82. 
Stars in the disk of M82 are traced out to major axis distances of 12 kpc, which is 
comparable to the spatial extents of stellar disks in other nearby late-type spirals. 
The number of bright AGB stars per unit surface brightness (the `specific frequency' -- SF) 
in the outer disks of M82 and the Sc galaxy NGC 2403 are similar. To the extent that red 
and infrared surface brightness serve as a proxy for projected stellar mass density, then this 
suggests that the specific SFR in the outer disks of these galaxies were similar during 
intermediate epochs. That the radial size of the stellar disk and the 
SF of bright AGB stars in M82 are similar to those in NGC 2403 
is consistent with M82 having been a gas-rich late-type disk galaxy 
prior to encountering M81. However, the SF of RSGs 
in the outer disk of M82 is much lower than in NGC 2403, 
suggesting that the star-forming histories of these galaxies diverged 
within the past $\sim 0.1$ Gyr. This is probably a consequence 
of the interaction with M81, which disrupted the ISM of M82.

	The interaction with M81 has also had a major impact on the extraplanar regions 
of M82. Feedback from the star-forming regions ejects material out of the disk plane, 
while the tidal interactions with M81 may also have pulled some of the ISM from the 
disk of M82, and deposited it around the galaxy. Given the presence of extraplanar gas, 
it is interesting that a mixture of bright main sequence stars, RSGs, 
and AGB stars that extend up to projected distances of at least $\sim 7$ kpc above the disk 
plane are seen in the CFHT data. The exponential scale height of bright AGB stars 
is $1.8 \pm 0.2$ kpc, which is considerably larger than what is traditionally 
associated with disk thickness. The photometric properties of 
the most evolved extraplanar RSG and AGB stars are consistent with 
them being at least moderately metal-rich, indicating that they formed from material 
that likely came from a disk, as opposed to a chemically immature 
environment. There are also five objects that have photometric 
and structural properties that are consistent with them being metal-poor globular 
clusters. Given that (1) stars evolving on the RSG and AGB sequences have different ages, 
and (2) the progenitor of M82 presumably contained a modestly populated classical 
(e.g. old and metal-poor) halo, of which globular clusters are the brightest 
tracers, then the extraplanar regions of M82 contain stars spanning a broad range of 
ages and metallicities. 

\subsection{A Spatially Extended Stellar Disk in M82}

	Individual bright AGB stars have been detected in the CFHT images out to 12 kpc 
along the major axis of M82. The SF of AGB stars between major axis distances 
of 4 and 12 kpc is constant, indicating that the AGB stars (1) belong to the 
disk, and (2) are uniformly mixed throughout the disk. Therefore, the stellar disk of 
M82 extends to much larger radii than sampled by, for example, 2MASS data, 
which traces the exponential light profile of M82 out to 350 arcsec ($\sim 6.6$ kpc) 
(Jarrett et al. 2003).

	The stellar disk of M82 almost certainly extends to distances $> 12$ kpc. 
Unfortunately, lacking the spectroscopic information that would be helpful to distinguish 
between stars and galaxies, contamination from background galaxies frustrates 
efforts to push the detection of AGB stars to larger radii. 
This being said, the linear size of the M82 stellar disk as mapped with the CFHT data is 
at the small end of what is seen in other nearby late-type spirals that are at comparable 
distances and have been investigated using the same instrumentation. The stellar disk of 
NGC 2403 has been traced out to $\sim 14$ kpc (Davidge 2007), while in NGC 247 the stellar 
disk extends to at least $\sim 18$ kpc (Davidge 2006). However, 
when gauged in terms of disk scale length, the M82 stellar disk 
is relatively large. With an exponential scale length of $\sim 0.9$ kpc 
(Mayya et al. 2005), then AGB stars in M82 are seen out to 13 scale lengths. For 
comparison, stars in NGC 247 and NGC 2403 are traced out to 
$\sim 7$ scale lengths (Davidge 2006; 2007).

	Mayya et al. (2005) trace spiral structure in near-infrared images of 
M82 out to $\sim 2$ arcmin, or $\sim 3$ kpc. Even though the stellar 
disk of M82 extends out to much larger radii (e.g. \S 3.1), 
spiral structure may not be expected in stellar light at 
distances larger than that found by Mayya et al (2005). 
The random motions imparted to stars as they 
encounter other objects in the disk cause them to diffuse from their places of birth, 
and spiral structure becomes increasingly blurred as the spatial 
distribution of stars with progressively larger ages are considered. Davidge (2007) 
found that the brightest main sequence stars in NGC 2403, 
which have ages $\leq 10$ Myr, are excellent tracers of spiral structure, 
whereas RSGs with ages $\sim 50$ Myr are not. Given that 
stars with ages of $\sim 10$ Myr are rare outside of the central few kpc of 
M82 and that the brightest main sequence stars in the outer 
disk likely have ages approaching 100 Myr, then stellar light will not trace 
spiral arms outside of the inner few kpc of M82.

	The time since the interaction 
between M81 and M82 is comparable to the crossing time of the M82 disk, 
and so large-scale signatures of tidal disturbances in the stellar 
disk might still be evident if they were ever present, and none are seen. 
That the disk of M82 extends out to many scale lengths and 
that the distribution of AGB stars does not show warping out of the disk plane
suggests that the outer stellar disk was not greatly disrupted. 
This is perhaps surprising given that 
Yun, Ho, \& Lo (1993) find that the M82 gas disk is warped, 
while Yun et al. (1994) conclude that M82 likely harboured a 
large HI disk prior to interacting with M81, but that much of this gas has now been lost. 
There are a number of probable tidal structures near M81 and its companions, and 
at least some of these may have formed from gas that originated in M82.
Indeed, simulations suggest that it is gas, rather than stars or dark 
matter, that is the key ingredient for the formation of long-lived tidal structures 
(Wetzstein, Naab, \& Burkert 2007). 

	The large-scale behaviour of gas in and around the M82 disk is broadly 
consistent with model predictions. Brouillet et al. (1992) modelled the
interactions between M81, M82, and NGC 3077, and reproduced the gaseous features that link 
these galaxies. However, these models assume that the disk of M82 was not greatly disturbed, 
whereas the distribution of HI and molecular material at the present day indicates that 
this is not the case. Still, the depleted ISM in M82 is concordant with 
models that demonstrate the fragility of disks 
during galaxy-galaxy encounters (e.g. Barnes 1992). Struck (1997) 
investigated models in which galaxies of unequal size, consisting of gas disks and 
rigid halos, collide. These simulations predict that the gas disks of both 
galaxies are drastically affected. Tidal forces cause 
the disk of the larger galaxy to expand, and a gas ring may form around this galaxy. 
A gas bridge, consisting mainly of material 
from the smaller galaxy, connects the two galaxies, 
while the gas that remains in the smaller galaxy contracts in size. 
The contraction of the gas disk of the smaller galaxy is consistent with 
the evidence that tidal interactions trigger the inward flow of disk gas 
(e.g. Iono et al. 2004; Kewley et al. 2006), which fuels centrally 
concentrated star-forming activity.

\subsection{The Stellar Content of the Outer Disk}

\subsubsection{A Recent Decline in the SFR in the Outer Disk of M82}

	One motivation for studying nearby galaxies is that it is possible to resolve 
individual stars in areas that have also been studied spectroscopically. 
Mayya et al (2006) found that the spectrum of the M82 disk in the 1 -- 3 kpc 
interval could be matched by that of a simple stellar system with an age $\sim 0.5$ 
Gyr. Ostensibly, this might suggest the star-forming history of the M82 disk during 
intermediate epochs was spectacular. However, this is a luminosity-weighted age, that is 
skewed by the low M/L ratios of intermediate age stars. In \S 3.3 it 
is shown that the SF of AGB stars with ages $\sim 0.5$ Gyr at 
large radii in M82 is similar to that in the outer regions of 
NGC 2403. The Sc galaxy NGC 2403 is relatively isolated, and is 
well removed from M81 and its companions (Karachentsev et al 2002). Thus, it 
is reasonable to assume that it has evolved in comparative isolation. 
To the extent that NGC 2403 is a representative `normal' late-type spiral galaxy, 
then this purely empirical comparison suggests that the specific SFR in M82 
$\sim 0.5$ Gyr in the past was not markedly different from that in other late-type 
spiral galaxies. Given that the SFR in galaxy disks is likely to be 
roughly constant with time, then spectra of these objects 
should contain Balmer absorption features that are consistent with a population that 
formed within the past 1 Gyr that comprises $\sim 10\%$ by mass of the stellar content. 
The visible light spectrum will then be that of an intermediate age population (e.g. Serra \& 
Trager 2007), in agreement with that observed by Mayya et al. (2006) in M82. 

	The comparisons with NGC 2403 indicate that 
the outer disk of M82 is deficient in bright RSGs and main sequence stars. 
The relative paucity of RSGs in the outer disk of M82 
suggests that the SFR dropped with respect to that in NGC 2403 within the past 
$\sim 0.1$ Gyr. Given that other indicators suggest that M81 and M82 interacted at least 
0.2 Gyr in the past, then it appears that star formation continued in the outer disk of M82 
for a significant period of time after the interaction. Still, 
given the evidence that the ISM of M82 was affected by the interaction with M81 
then it seems reasonable to conclude that the decline 
in the SFR 0.1 Gyr in the past is probably somehow linked to the interaction with M81.

	The reader is cautioned that as highly evolved, relatively rare objects, 
RSGs are not an optimum probe of star-forming history. Observations of stars at 
the main sequence turn-off (MSTO) in the outer disk of M82 will provide a 
more reliable chronometer for determining when the SFR in the outer disk of M82 declined, 
and so the 0.1 Gyr age estimated for the drop in the SFR should be considered to be 
preliminary. The MSTO in the outer disk of M82 will occur near M$_V \sim -2$ (i.e. $V 
\sim 26$) if the SFR dropped 0.1 Gyr in the past. 

\subsubsection{Sources of Uncertainty in the Age Scale}

	There are various sources of uncertainty in the ages estimated in this paper. 
Uncertainties in the metallicities of the stars
being investigated affects the age estimates. In \S 3 it was shown that the ages 
estimated for AGB stars from the Z = 0.008 and Z = 0.019 isochrones differ by $\sim 
0.5$ dex. Internal reddening is another possible source of uncertainty. It is likely 
that most of the dust at the present day is concentrated in the star 
forming regions near the center of M82. This being said, if 
the mean internal extinction measured for the central regions of the galaxy holds 
for the entire disk then the impact on ages is not great as long as the distance to 
M82 computed from RGB-tip measurements is also corrected for this reddening. 
Finally, the use of AGB stars as chronometers introduces 
uncertainties due to (1) the complicated model physics of highly evolved stars, 
(2) photometric variablity, and (3) stochastic effects due to the short evolutionary 
timescales of stars near the AGB-tip. The first of these can be mitigated somewhat by making 
direct comparisons between galaxies, as in \S 3.

	Crowding also complicates efforts to probe the brightest stars in nearby galaxies, 
as blends of faint stars may appear as objects that skew 
age estimates to younger values. To reduce the impact of crowding, 
only stars that (1) are well outside of the crowded central regions of the galaxy 
and (2) are at brightnesses that the artificial star experiments indicate are 
not significantly contaminated by blends, have been considered in this study. The excellent 
annulus-to-annulus agreement in the SF measurements 
suggests strongely that crowding is not an issue when R$_{GC} > 4$ kpc, and this is 
consistent with predictions from artificial star experiments. 

	A comparison of the properties of AGB stars from the 
MegaCam and WIRCam datasets can be used to test the claim that blending 
is not significant among the brightest AGB stars. This is because crowding has a very 
different impact on the brightest red stars at visible and near-infrared wavelengths, 
due largely to the wavelength-dependence of line blanketing. As the impact of line 
blanketing drops with increasing wavelength, there is increased contrast 
between red AGB stars and stars on the RGB, with the result that 
at infrared wavelengths there is a lower fraction of blends between RGB stars 
that may masquerade as bright AGB stars than at visible wavelengths.

	If blending significantly affects the MegaCam observations more than the WIRCam 
observations then the M$_{bol}$ LFs of AGB stars generated from the MegaCam data 
should contain stars that are brighter than those derived from the WIRCam data.
To compare the LFs of the two datasets, the magnitudes of stars that have 
photometric properties that are consistent with M giants in the $(i', r'-i')$ and $(K, 
J-K)$ CMDs were converted into bolometric magnitudes. The comparison is restricted to 
M giants, as these are the stars that are common to both the MegaCam and WIRCam data. 
Bessell \& Wood (1984) compute bolometric corrections (BCs) as functions of $R-I$ 
and $J-K$ colors, and these calibrations are adopted here. 
$R-I$ colors were computed from the $r'-i'$ colors using the transformation 
equations from Smith et al. (2002). Relations between T$_{eff}$ and $R-I$ and $J-K$ 
were defined from data tabulated by Bessell (1979) and Bessell \& Brett (1988).

	The M$_{bol}$ LFs of stars with log(T$_{eff}$) between 3.54 and 3.60 in two 
radial intervals are compared in Figure 17. The comparisons are affected 
by uncertainties in the BC and T$_{eff}$ calibrations, and a large fraction 
of highly evolved AGB stars are expected to be photometrically variable. Still, the LFs 
constructed from the MegaCam and WIRCam datasets are in reasonable agreement, 
suggesting that blending does not affect the bright AGB stellar content 
in the $(i', r'-i')$ CMDs.

\subsection{Young and Intermediate-Age Stars in the Extraplanar Regions of M82}

	Stars with ages log(t) $\leq 9.0$ have been detected out to projected distances 
of $\sim 7$ kpc above the M82 disk plane. Where did these stars form? The 
origin of these stars is discussed in this section.

	Seth, Dalcanton, \& de Jong (2005) investigate the vertical distribution of 
stars in a sample of nearby edge-on low mass disk galaxies. The vertical height of a star 
has an age dependence, in the sense that young stars are found closer to the disk 
than older stars. Seth et al. (2005) suggest that these stars belong to thick disks, 
that are populated by stars that leave the disk plane through 
dynamical heating, such as encounters with molecular clouds. 
Stars are detected up to 3.5 kpc above the disk plane, and Seth et al. conclude 
that the vertical scale height of thick disks in pure disk systems is 
larger than in the Galaxy. The young stars found in M82 extend to much larger 
distances off of the disk plane than those considered by Seth et al. (2005). 
Blue main sequence stars are also seen throughout the extraplanar regions of M82, 
whereas they should hug the disk plane if they were thick disk objects. While these arguments 
suggest that the majority of objects found in the extraplanar environment of M82 do not belong 
to a thick disk component, some stars in the $D_Z = 2.5 - 3.5$ kpc interval may be part 
of a thick disk like that seen in other disk systems.

	A modest fraction of young disk stars will acquire high velocities due to 
dynamical processes. A fraction of OB stars in the Solar neighborhood have peculiar 
velocities $\geq 30$ km sec$^{-1}$, and if these stars have space motions directed 
out of the disk plane then they could populate at least part of the extraplanar regions 
in the Milky-Way. Various mechanisms have been proposed to explain 
the high velocities of these stars, including 
ejection from close binary systems due to SN explosions, and 
dynamical ejection from young open clusters. Kiseleva et al. (1998) 
estimate that $\sim 1\%$ of stars that form in small stellar systems will become 
high-velocity stars. 

	Could the youngest stars in the extraplanar regions of M82 be runaway objects? 
With a peculiar velocity of 30 km sec$^{-1}$ then a star with an age of $10^8$ years 
could traverse 3 kpc in its lifetime. Therefore, populating the extraplanar regions of M82 
with stars that are younger than $\sim 100$ Myr would require velocities in excess 
of 60 km sec$^{-1}$. Even if such stars were moving 
perpendicular to the disk plane, the maximum height that they could attain depends on the 
local gravitational potential, and stars ejected near the centers of galaxies at 
a given velocity will not reach the same distances above the disk as those 
in the outermost regions having the same velocity. While they may constitute some fraction 
of the young stellar component that is closest to the disk plane, stars ejected 
from the disk of M82 by dynamical processes probably can 
not populate the full extent of the extraplanar regions that have been probed in M82. 

	Given the evidence that the gas disk of M82 has been disrupted, then it is 
possible that some young and intermediate-age stars may have been pulled from the disk. 
If tidal forces populated the extraplanar regions of M82 
then deeper observations should reveal two characteristics of the stellar content. 
First, tidal forces would have pulled a mix of young and old stars from the disk, so 
there should be an RGB component with a disk-like (i.e. moderately high) metallicity. 
Second, the spatial distribution of the stars pulled from the disk should not depend on age, 
and so the metal-rich RGB stars should be distributed with an exponential scale height 
that is the same as that of intermediate age AGB stars, which is 1.8 kpc. 
A major problem with the tidal origin model is that extraplanar stars with ages $< 100$ 
Myr are hard to explain, as these objects formed well after M81 and M82 interacted.

	It seems likely that some of the young and intermediate age stars in the 
extraplanar regions of M82 formed {\it in situ}. There are obvious 
difficulties forming stars in a hot outflow. Still, the formation of a shock where 
the outflow encounters gas or dust clouds surrounding M82, such as is thought to be 
occuring in the Cap (e.g. Lehnert et al. Strickland et al. 2004), could cool hot gas so that 
star formation can occur. de Mello et al. (2008) find star-forming regions in Arp's Loop, 
which is an area of local HI concentration in the tidal debris trail between M81 and M82.

	As discussed by Davidge (2008), the stars associated 
with M82 South likely do not form a bound structure, and tidal forces will likely 
disperse them in $\sim 10^8 - 10^9$ years. de Mello et al. (2008) point out that 
the fate of the star-forming regions in Arp's Loop is less clear.
Still, if other structures like M82 South and those in the Arp Loop have 
formed in the past and have been disrupted then they may produce a 
diffusely-distributed population of youngish stars in the extraplanar regions of M82 and 
throughout the M81-M82 debris field. 

	The presence of relatively young stars may explain the extraplanar UV 
emission seen in M82. This emission is thought to come from 
UV radiation that is reflected from dust clouds. The source 
of the reflected light is not clear, although Hoopes et al. (2005) suggest that it may 
be the stellar continuum from the starburst. The FUV emission in the Hoopes et al. (2005) 
study is strongest along the minor axis to the south of the galaxy, and this is also 
where the largest number of candidate main sequence stars have been found (\S 4). Given 
that M82 South (1) is also a source of UV emission in the Hoopes et al. (2005) study, 
and (2) contains stars with ages of $\sim 50$ Myr (Davidge 2008), 
then at least some of the FUV emission may come from 
main sequence stars in the extraplanar regions. In the particular case of M82 South, 
the strip of emission at visible wavelengths that defines the southern boundary of the 
stellar distribution may be reflected light from main sequence stars in M82 South.

	We close by noting that the stellar content of the extraplanar regions of M82 
may provide clues for understanding the halos of other galaxies. It has been shown 
in this paper that stars in the extraplanar regions of 
M82 have diverse properties, with objects spanning ages from log(t) $\sim 7.5 - 9.0$. Older 
stars are almost certainly present, but these are too faint to detect with the CFHT data. 
There are hints of an old extraplanar component in the form of globular clusters, 
although even if subsequent observations confirm that they are globular clusters they may 
still belong to M81.

	Comparisons with isochrones indicate that the young and intermediate age 
stars in the extraplanar regions have a disk-like metallicity, which is higher than 
what might be associated with a classical halo. An 
observer a few Gyr in the future, well after the starburst 
activity has ceased, may see M82 as a disk galaxy with a marked spread in the metallicity 
and age of stars off of the disk plane. Such a dispersion in stellar content appears not 
to be rare in the outer regions of nearby spiral galaxies (Mouchine 2006), suggesting that 
interactions in the past may have had a role in populating the halos of many other galaxies. 

\acknowledgements{It is a pleasure to thank the anonymous referee for comments that 
resulted in a greatly improved paper.}

\parindent=0.0cm

\clearpage

\begin{table*}
\begin{center}
\begin{tabular}{cccccc}
\tableline\tableline
\# & RA & Dec & $K$ & $J-K$ & FWHM \\
 & (2000.0) & (2000.0) & & & (arcsec) \\
\tableline
1 & 09:57:18 & 69:40:30 & 19.9 & 0.92 & 0.6 \\
2 & 09:57:11 & 69:26:18 & 19.1 & 0.45 & 0.7 \\
3 & 09:56:42 & 69:26:11 & 18.3 & 0.82 & 0.5 \\
4 & 09:56:13 & 69:25:00 & 17.4 & 0.83 & 0.4 \\
5 & 09:54:48 & 69:36:16 & 20.1 & 0.75 & 0.6 \\
\tableline
\end{tabular}
\end{center}
\caption{Globular Cluster Candidates}
\end{table*}

\clearpage
\begin{figure}
\figurenum{1}
\epsscale{0.85}
\plotone{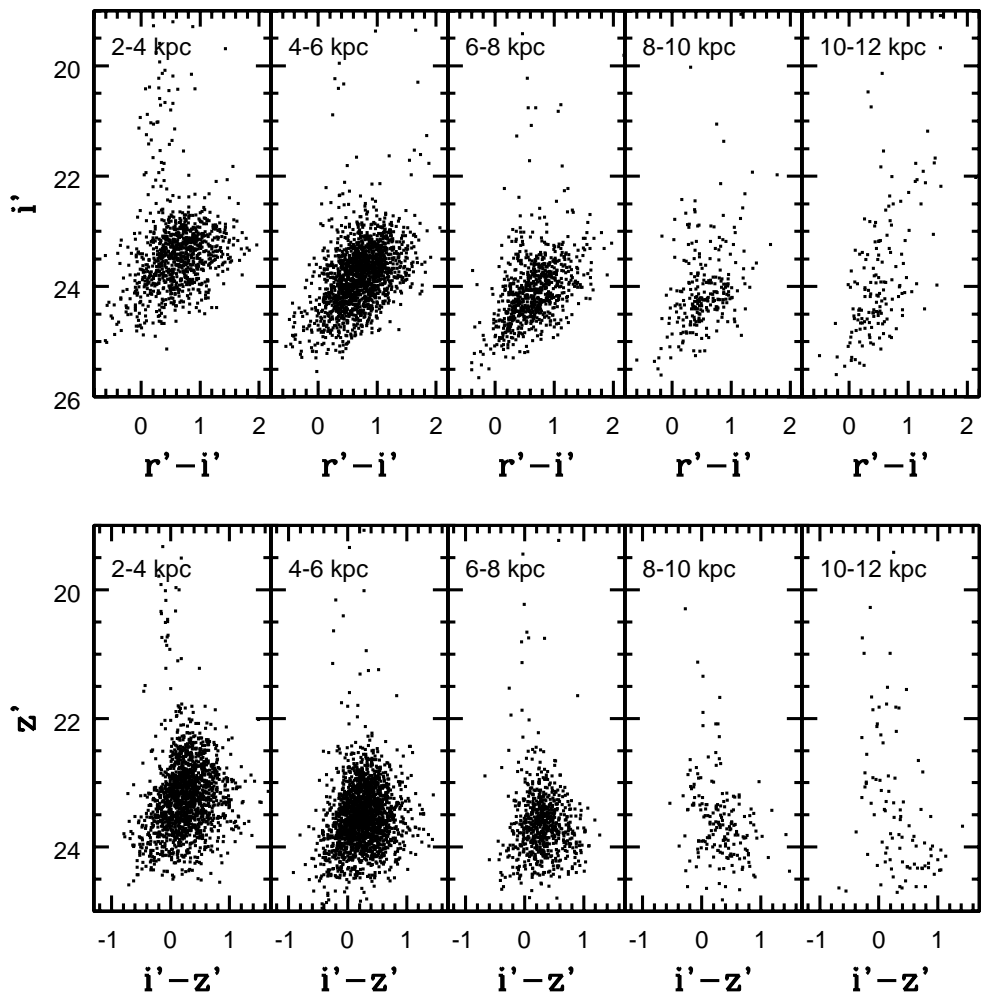}
\caption
{The $(i', r'-i')$ and $(z', i'-z')$ CMDs of stars within $\pm 65$ arcsec 
of the major axis of M82. The concentration of objects in the lower half of the CMDs 
is made up of bright AGB stars in M82. These objects can be traced out to at least 
R$_{GC} \sim 12$ kpc in M82.}
\end{figure}

\clearpage
\begin{figure}
\figurenum{2}
\epsscale{0.85}
\plotone{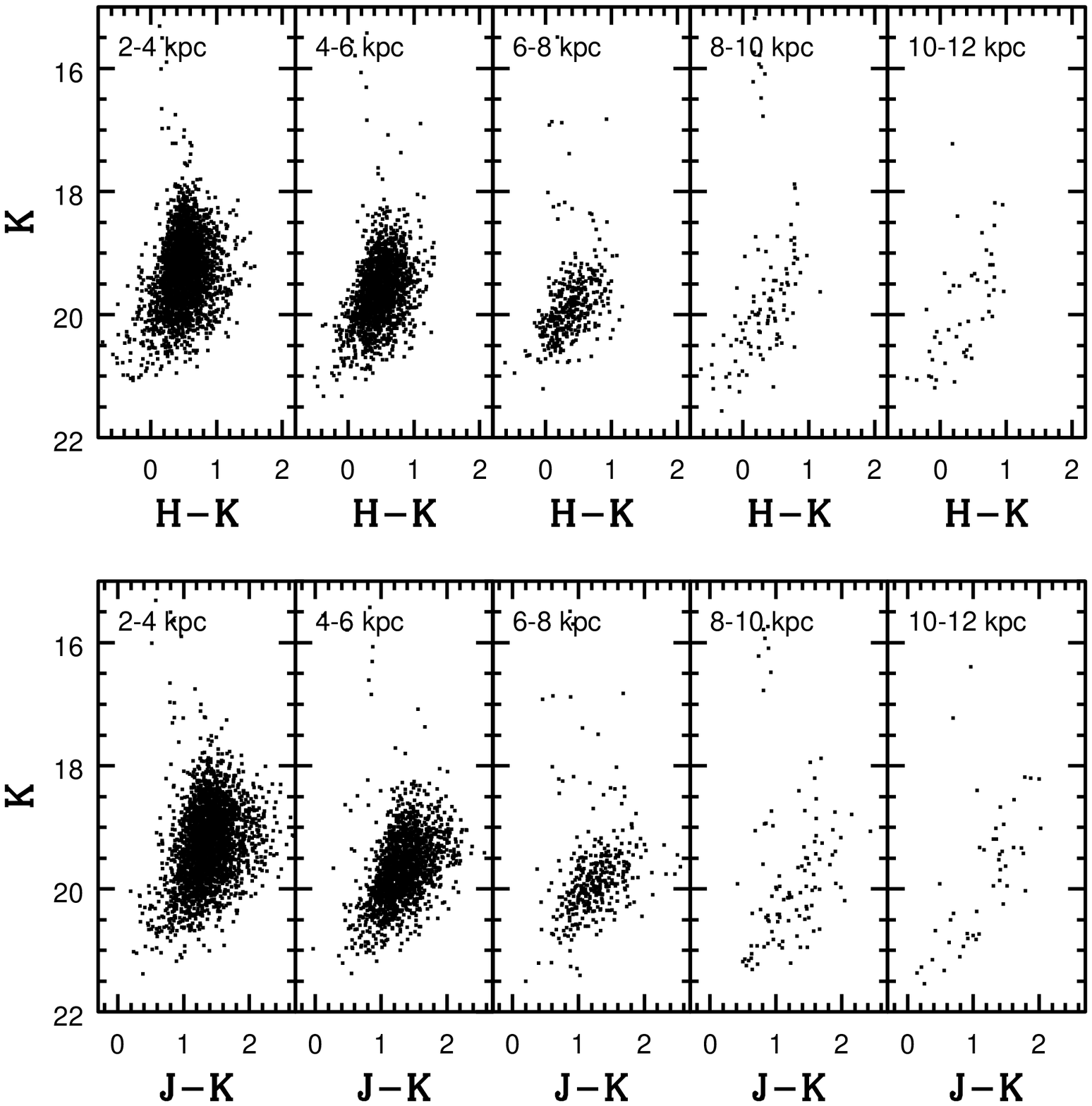}
\caption
{The same as Figure 1, but showing the $(K, H-K)$ and $(K, J-K)$ CMDs.}
\end{figure}

\clearpage
\begin{figure}
\figurenum{3}
\epsscale{0.85}
\plotone{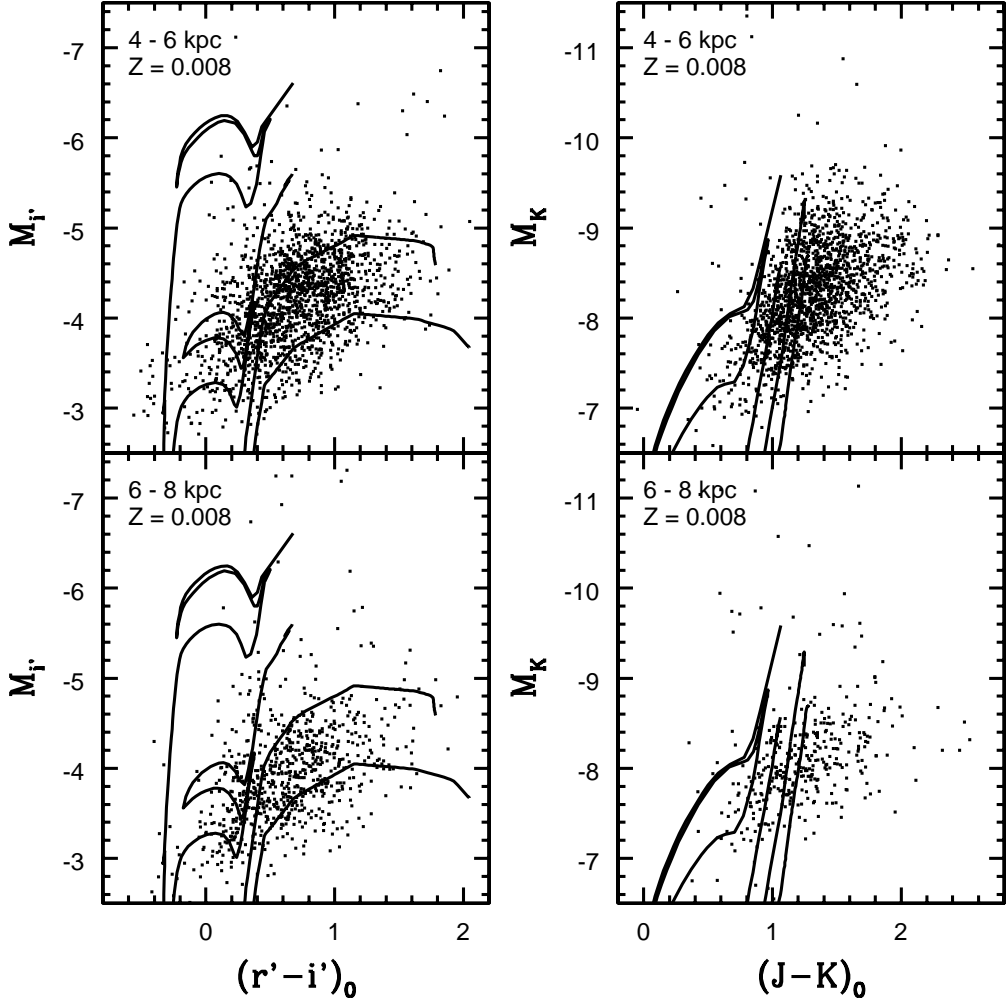}
\caption
{The CMDs of the 4 -- 6 kpc and 6 -- 8 kpc intervals 
in the M82 disk are compared with Z = 0.008 isochrones from Girardi 
et al. (2002; 2004). The isochrones have 
log($t_{yr}$) = 7.5, 8.0, 8.5, and 9.0. A distance modulus 
$\mu_0 = 27.95$ has been assumed, with A$_B = 0.12$ magnitude.}
\end{figure}

\clearpage
\begin{figure}
\figurenum{4}
\epsscale{0.85}
\plotone{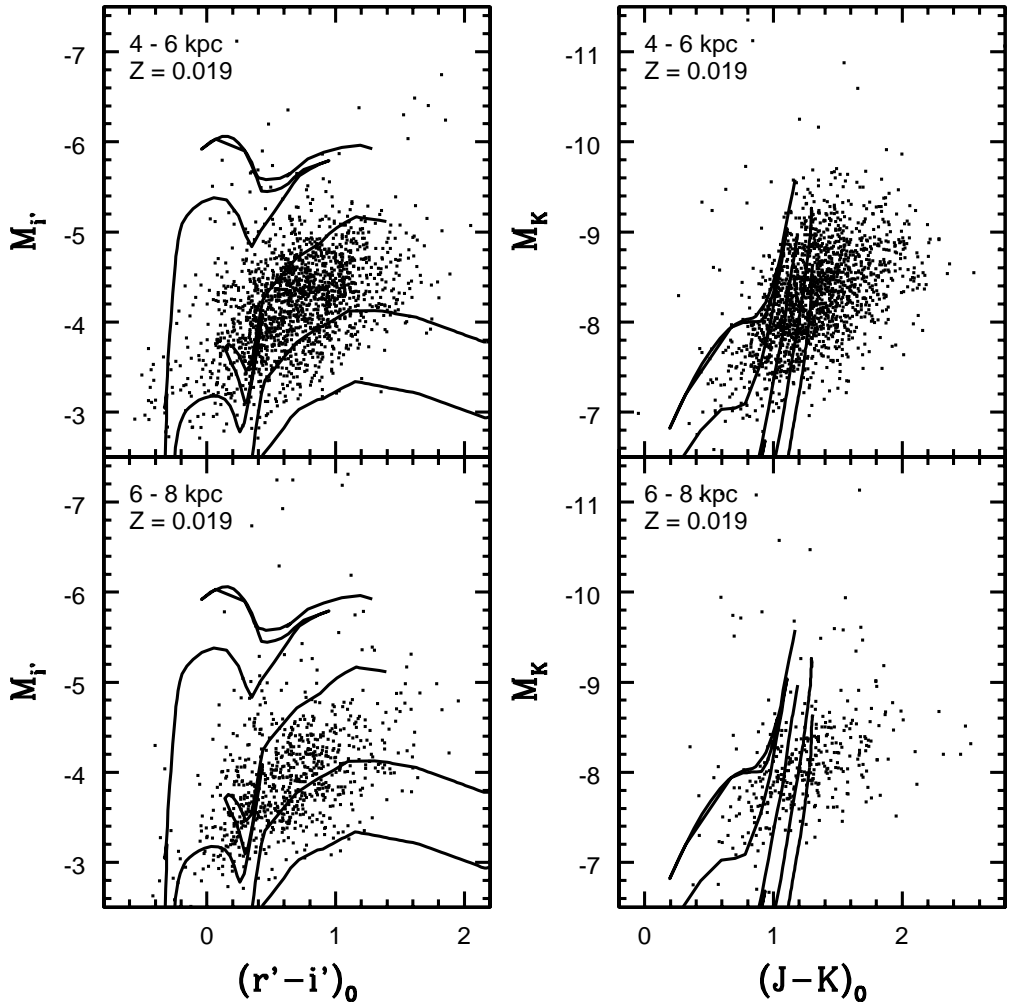}
\caption
{The same as Figure 3, but with Z = 0.019 isochrones.}
\end{figure}

\clearpage
\begin{figure}
\figurenum{5}
\epsscale{0.85}
\plotone{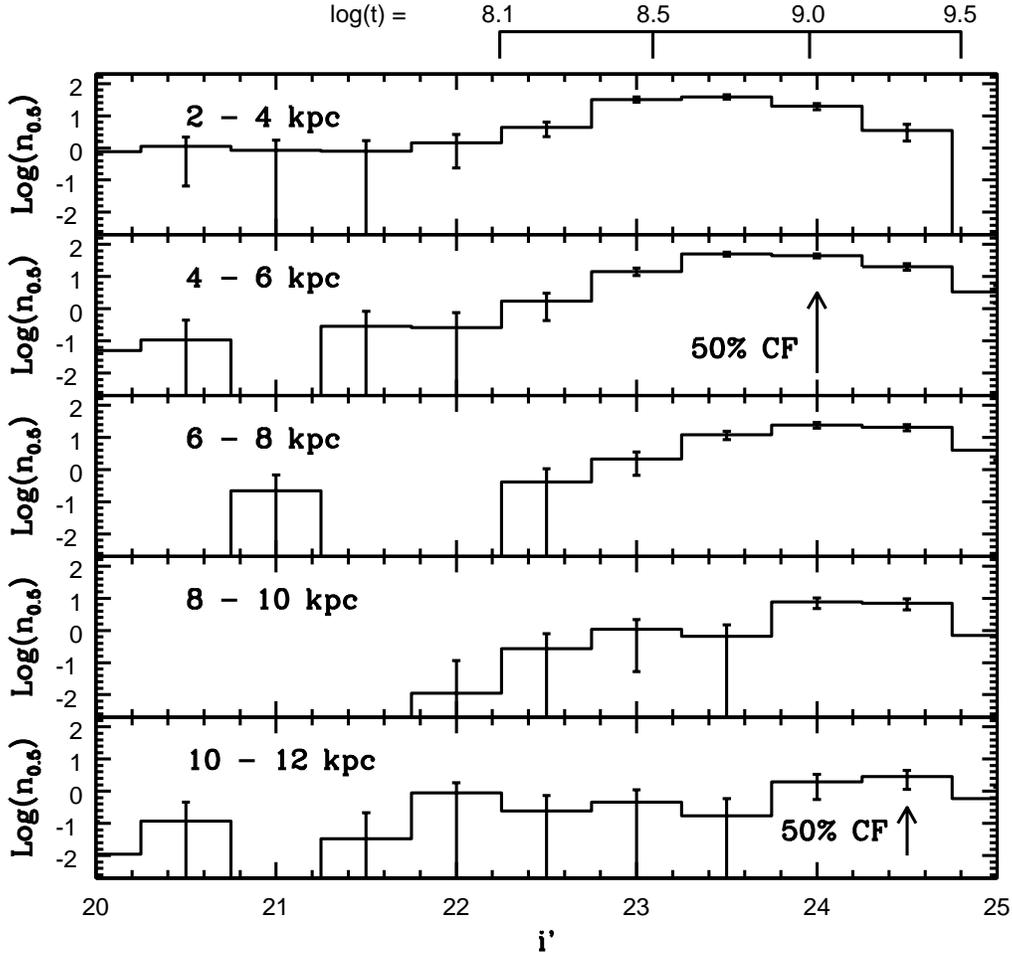}
\caption
{The $i'$ LFs of stars in the M82 disk. n$_{0.5}$ is the number of stars with $r'-i'$ 
between 0 and 2 per 0.5 $i'$ magnitude per arcmin$^{2}$. The LFs have been corrected for 
contamination from foreground stars and background galaxies by subtracting the LF of  
control fields, scaled to account for the areal coverage in each radial interval. 
The magnitude at which the star counts have a 50\% completeness fraction (50\% CF) is 
indicated for the 4 -- 6 and 10 -- 12 kpc intervals. The peak AGB brightnesses predicted 
by the Z = 0.008 Girardi et al. (2004) models for various ages are shown at the top of the 
figure.}
\end{figure}

\clearpage
\begin{figure}
\figurenum{6}
\epsscale{0.85}
\plotone{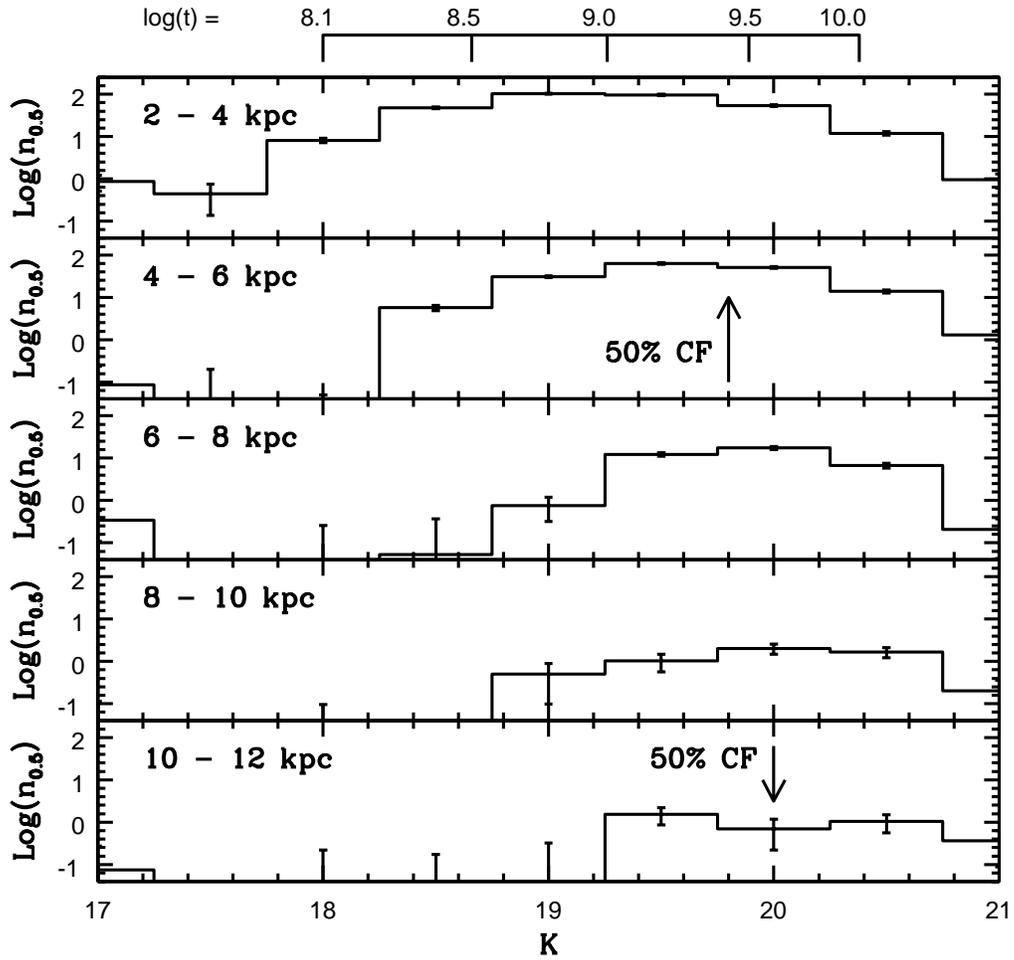}
\caption
{The same as Figure 5, but showing $K$ LFs. The age calibration is based on 
the peak AGB brightnesses in the Z = 0.008 Girardi et al. (2002) models.}
\end{figure}

\clearpage
\begin{figure}
\figurenum{7}
\epsscale{0.85}
\plotone{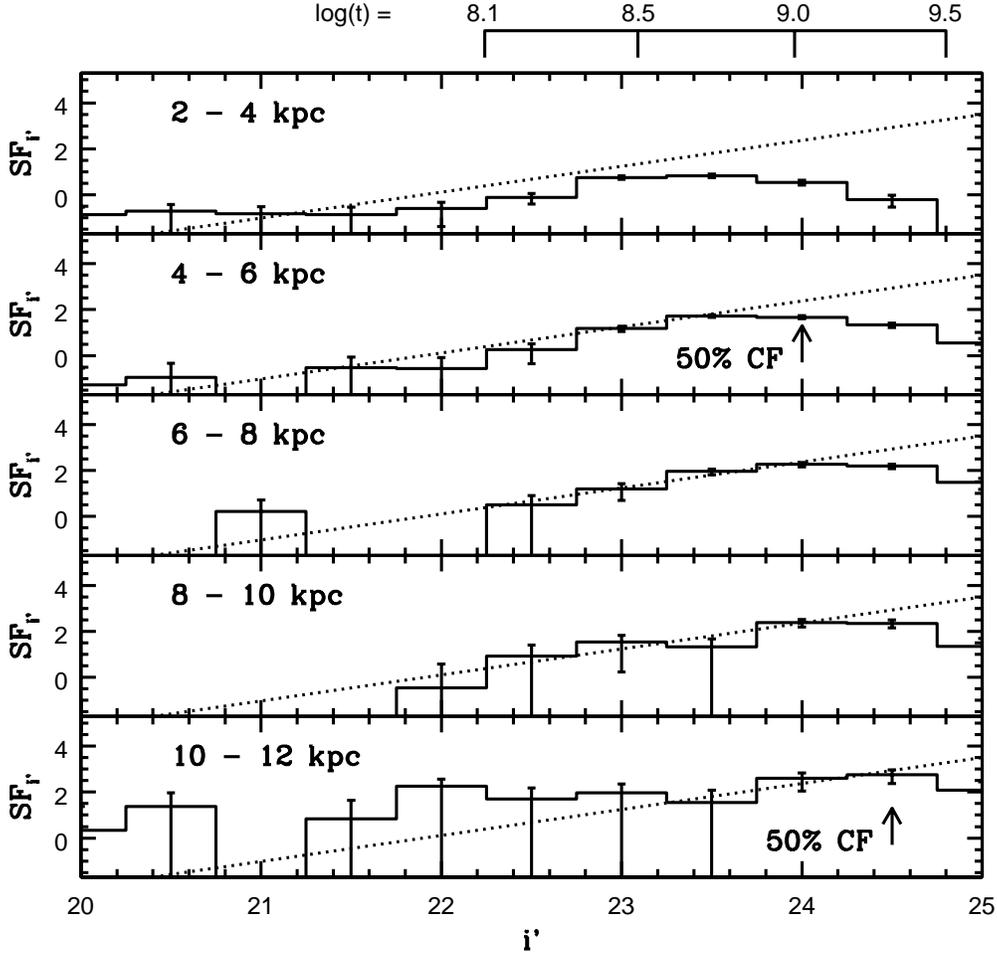}
\caption
{The specific frequency (SF) of stars measured from the MegaCam data. 
SF$_{i'}$ is the number of stars with $r'-i'$ between 0 and 2 per 0.5 $i'$ mag, scaled 
to match that in a system with a total magnitude M$_K = -16$. The dashed line shows the mean 
SF$_{i'}$ relation at intermediate brightnesses for stars with R$_{GC}$ between 4 and 10 kpc.
The magnitude at which the star counts have a 50\% completeness fraction 
(50\% CF) is indicated for the 4 -- 6 and 10 -- 12 kpc intervals. The peak AGB brightnesses predicted by 
the Z = 0.008 Girardi et al. (2004) models are shown at the top of the figure.}
\end{figure}

\clearpage
\begin{figure}
\figurenum{8}
\epsscale{0.85}
\plotone{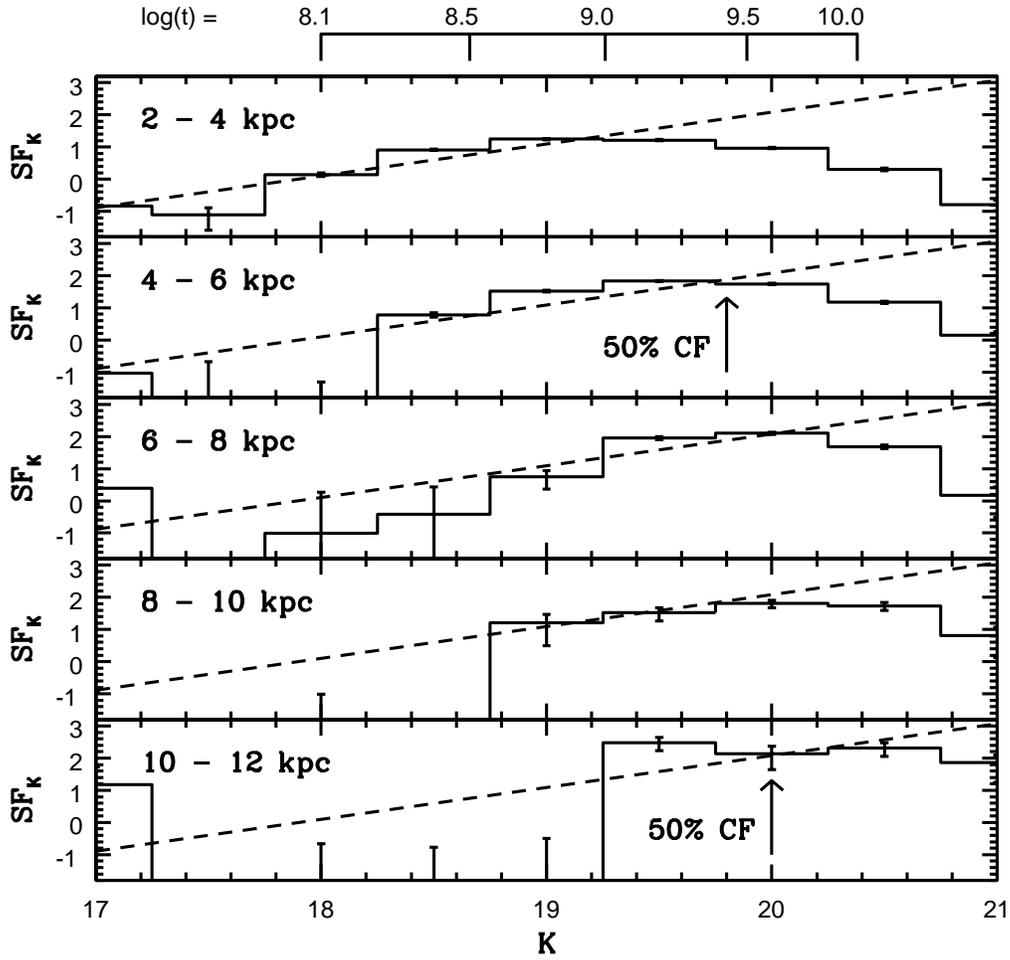}
\caption
{The same as Figure 7, but showing the SF of stars measured from the WIRCam data. 
SF$_{K}$ is the number of stars with $H-K$ between 0 and 1 per 0.5 $K$ mag, scaled to 
a system with a total integrated magnitude M$_K = -16$.}
\end{figure}

\clearpage
\begin{figure}
\figurenum{9}
\epsscale{0.85}
\plotone{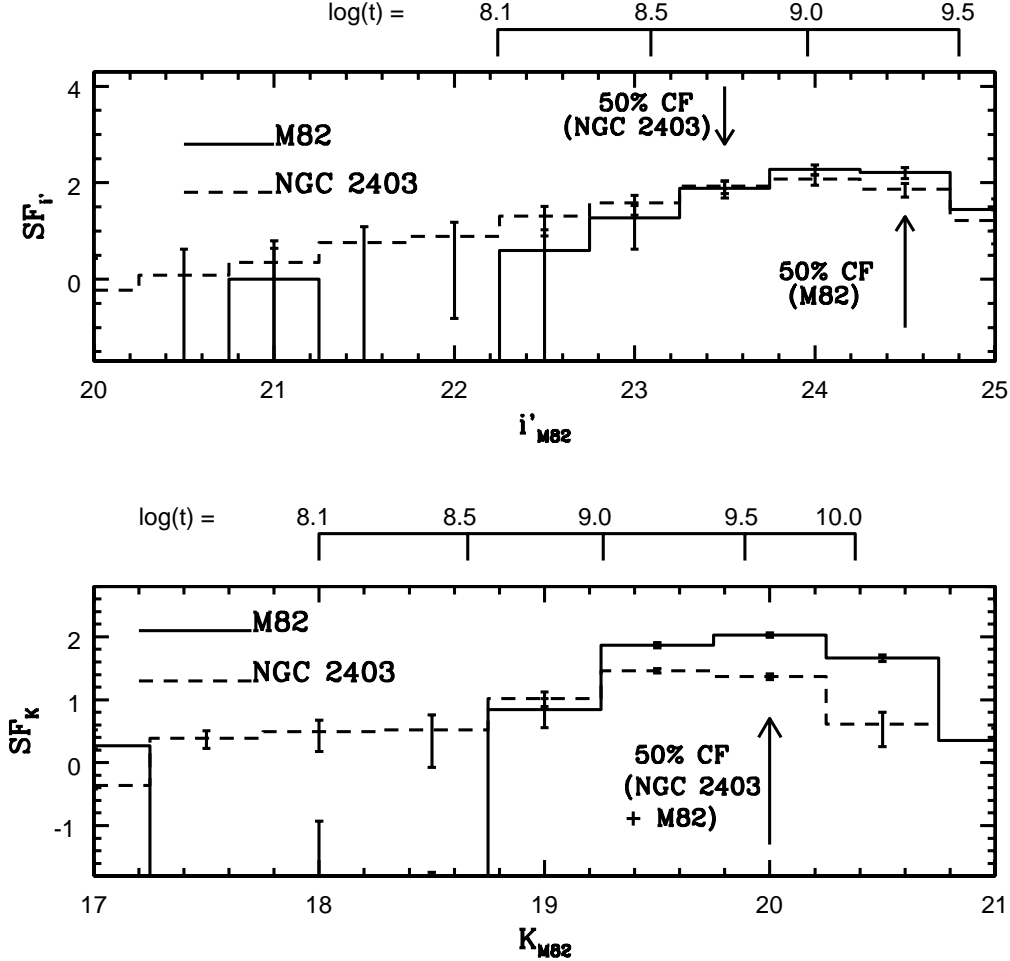}
\caption
{The SFs of stars in the outer disks of M82 (solid lines) and NGC 2403 
(dashed lines). SF$_{i'}$ and SF$_{K}$ are defined in Figures 7 and 8. 
The M82 SF measurements use stars with R$_{GC}$ between 6 and 10 kpc, while 
the NGC 2403 measurements include stars with R$_{GC}$ between 6 and 12 kpc. $i'_{M82}$ and 
$K_{M82}$ are the brightnesses that would be measured at the distance modulus of M82, and 
the age calibration at the top of each panel shows the peak AGB 
brightnesses from the Z = 0.008 Girardi et al. (2002; 2004) models. 
The magnitude at which the completeness fraction is 50\% 
(50\% CF) is indicated for each galaxy. Note the 
excellent agreement between the SF$_{i'}$ curves of the two galaxies 
at brightnesses that correspond to the AGB-tip of systems with ages $0.3 - 0.6$ Gyr. 
The SF$_{K}$'s also agree at $K = 19$, which corresponds to 
the AGB-tip brightness of a system with an age $\sim 0.6$ Gyr. The comparisons in the top 
panel also suggests that the SF of RSGs in NGC 2403 is higher than in M82.}
\end{figure}

\clearpage
\begin{figure}
\figurenum{10}
\epsscale{0.85}
\plotone{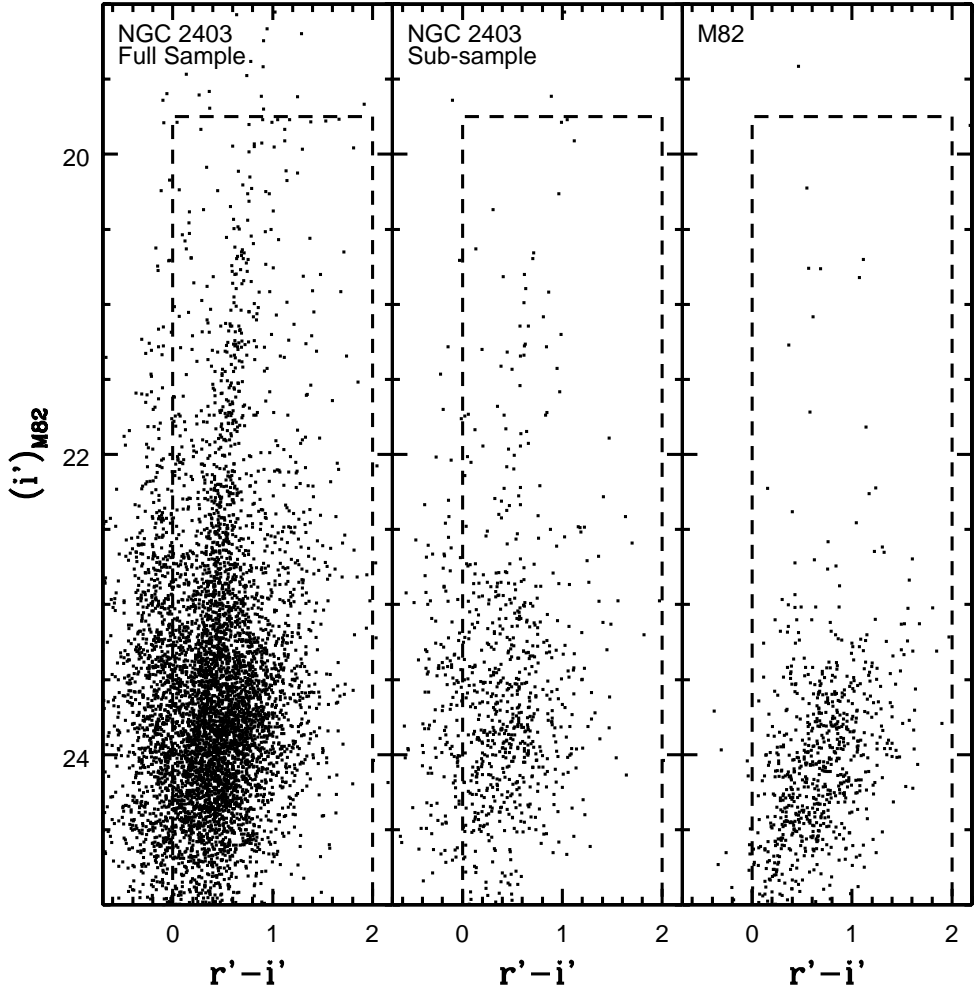}
\caption{The $(i', r'-i')$ CMDs of stars with R$_{GC}$ between 6 and 8 kpc in NGC 2403 
and M82. $i'_{M82}$ is the brightness that a star would have if observed at the same 
distance as M82. The left and right hand panels show the CMDs of all stars in the 6 - 8 kpc 
intervals of both galaxies. The central panel shows the CMD of a sub-sample of stars from 
the left hand panel that are located in small fields near the minor and major 
axes of NGC 2403 and that together contain the same number of stars with $i'_{M82}$ between 
23.25 and 23.75 as in the M82 CMD. The dashed lines indicate the color intervals that 
were used to generate the star counts shown in Figure 9. The CMDs of NGC 2403 and M82 are 
very different, in that the M82 CMD lacks (1) the plume of RSGs with $i' < 23$ and $r'-i'$ 
between 0.3 and 1.1 that is clearly seen in the NGC 2403 CMDs, and (2) 
the bright main sequence stars in NGC 2403 that have $r'-i' < 0$. 
The outer disk of M82 thus lacks the young population that is prevalent in the 
outer disk of NGC 2403, as expected based on the comparison in Figure 9.}
\end{figure}

\clearpage
\begin{figure}
\figurenum{11}
\epsscale{0.85}
\plotone{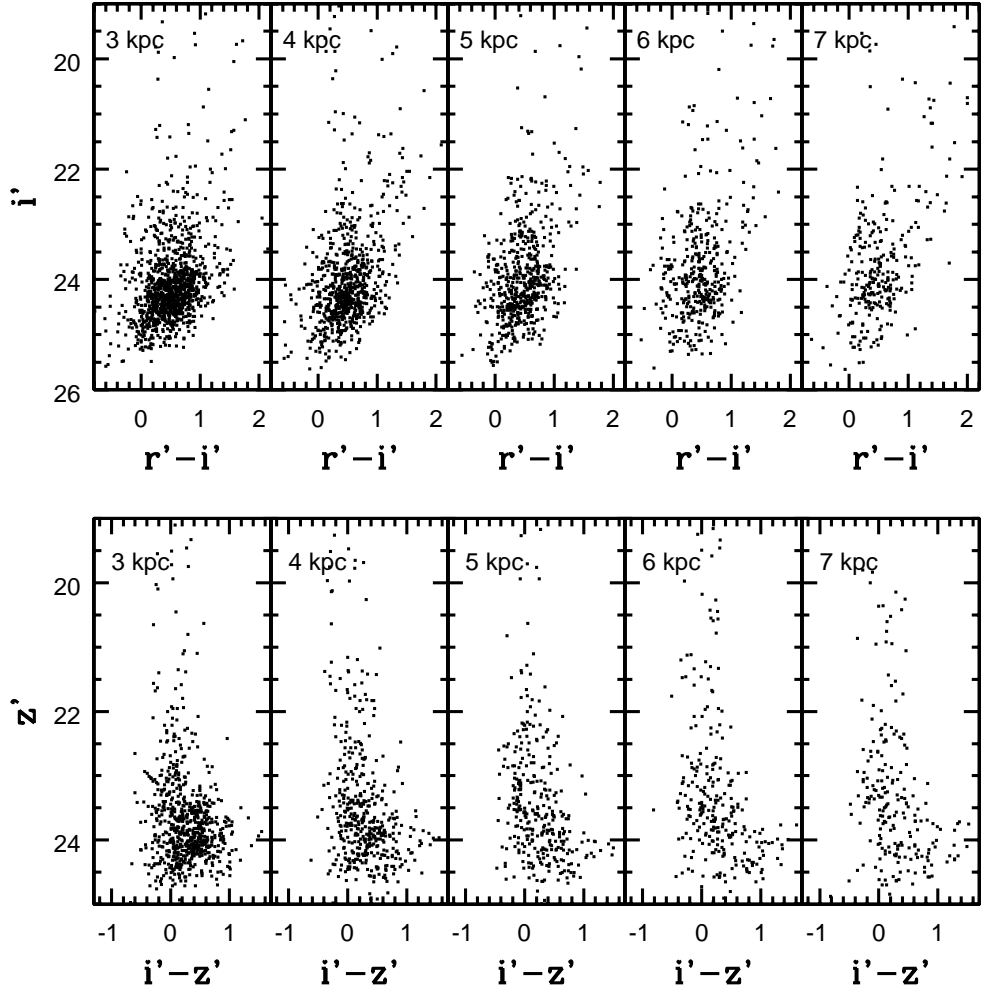}
\caption{The $(i', r'-i')$ and $(z', i'-z')$ CMDs of the 
extraplanar regions of M82. The distance given in each panel is 
measured from the center of M82 along the minor axis, assuming an ellipticity 
of 0.68 (Jarrett et as. 2003). Note that a concentration 
of AGB stars can be traced out to D$_Z = 7$ kpc.}
\end{figure}

\clearpage
\begin{figure}
\figurenum{12}
\epsscale{0.85}
\plotone{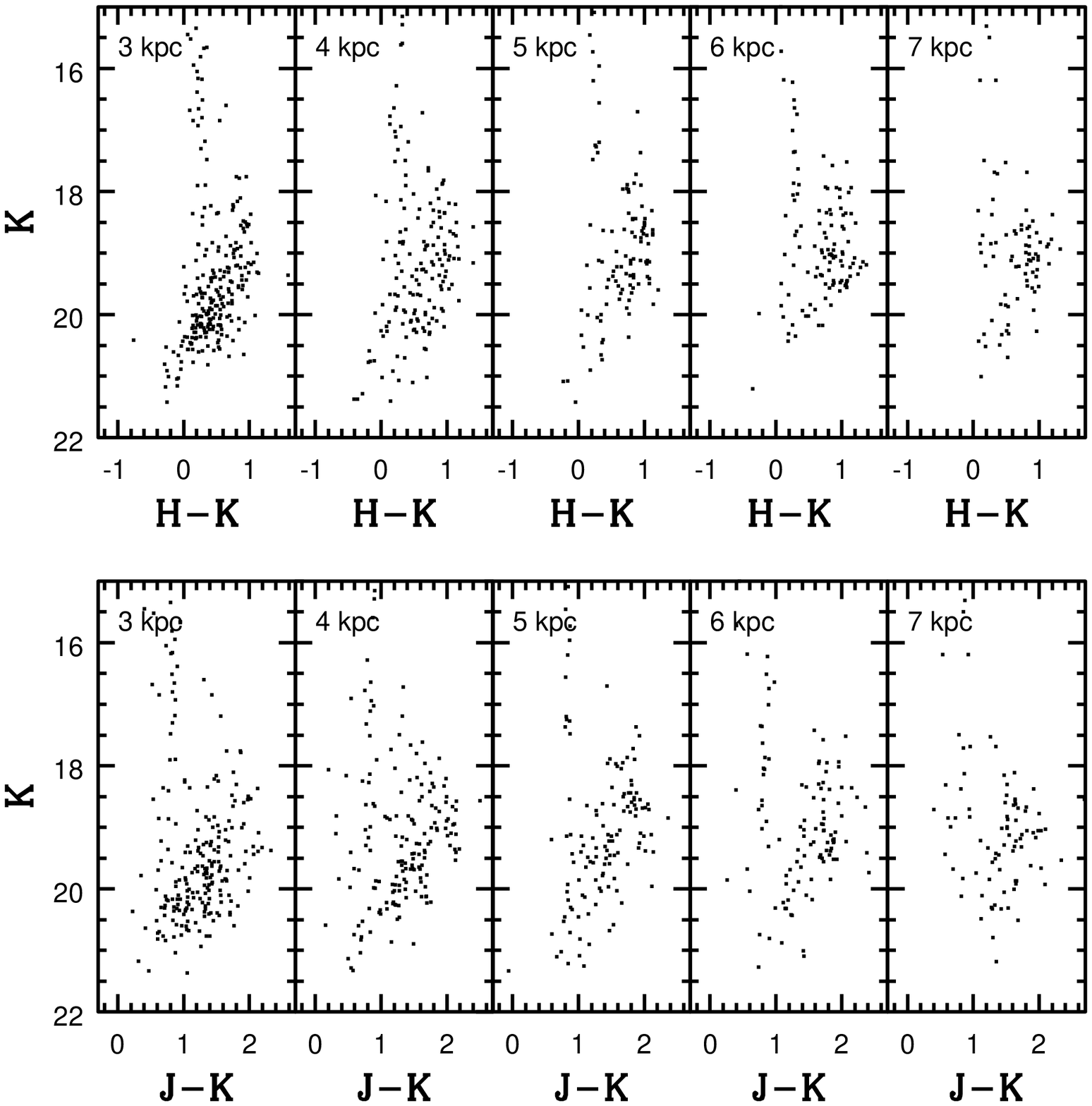}
\caption{The same as Figure 11, but showing the $(K, H-K)$ and $(K, J-K)$ CMDs.}
\end{figure}

\clearpage
\begin{figure}
\figurenum{13}
\epsscale{0.85}
\plotone{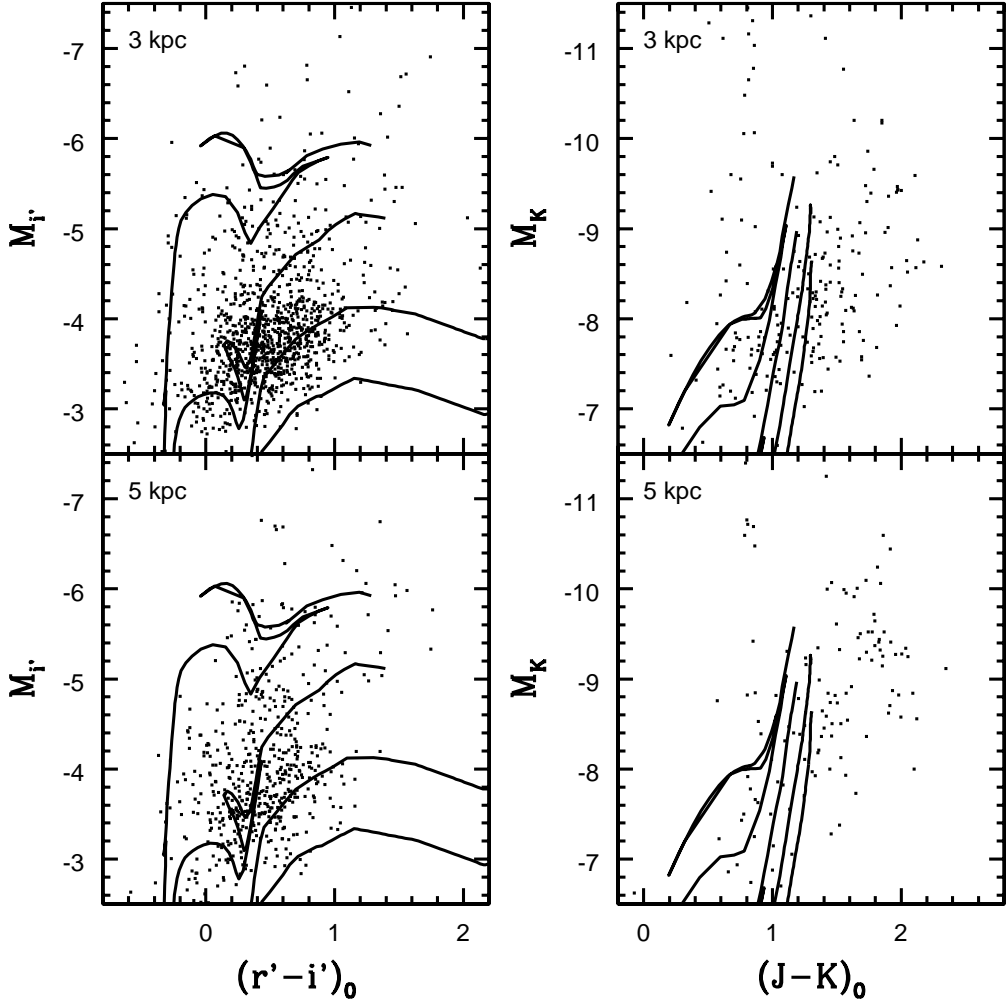}
\caption
{The (M$_{i'}, (r'-i')_0$) and (M$_K , (J-K)_0$) CMDs of the 
D$_Z = 3$ and 5 kpc intervals. The solid lines are Z = 0.019 isochrones from Girardi et 
al. (2002; 2004) with log(t$_{yr}$) = 7.5, 8.0, 8.5, and 9.0. Note that there are (1) blue 
objects in the (M$_{i'}, r'-i')$ CMDs that have colors and brightnesses that are consistent 
with main sequence stars having ages $< 0.1$ Gyr, and (2) 
sources that may be RSGs with ages $< 0.1$ Gyr. The objects with $(J-K)_0 > 1.6$ in the 
(M$_K, J-K)$ CMDs are probably background galaxies.}
\end{figure}

\clearpage
\begin{figure}
\figurenum{14}
\epsscale{0.85}
\plotone{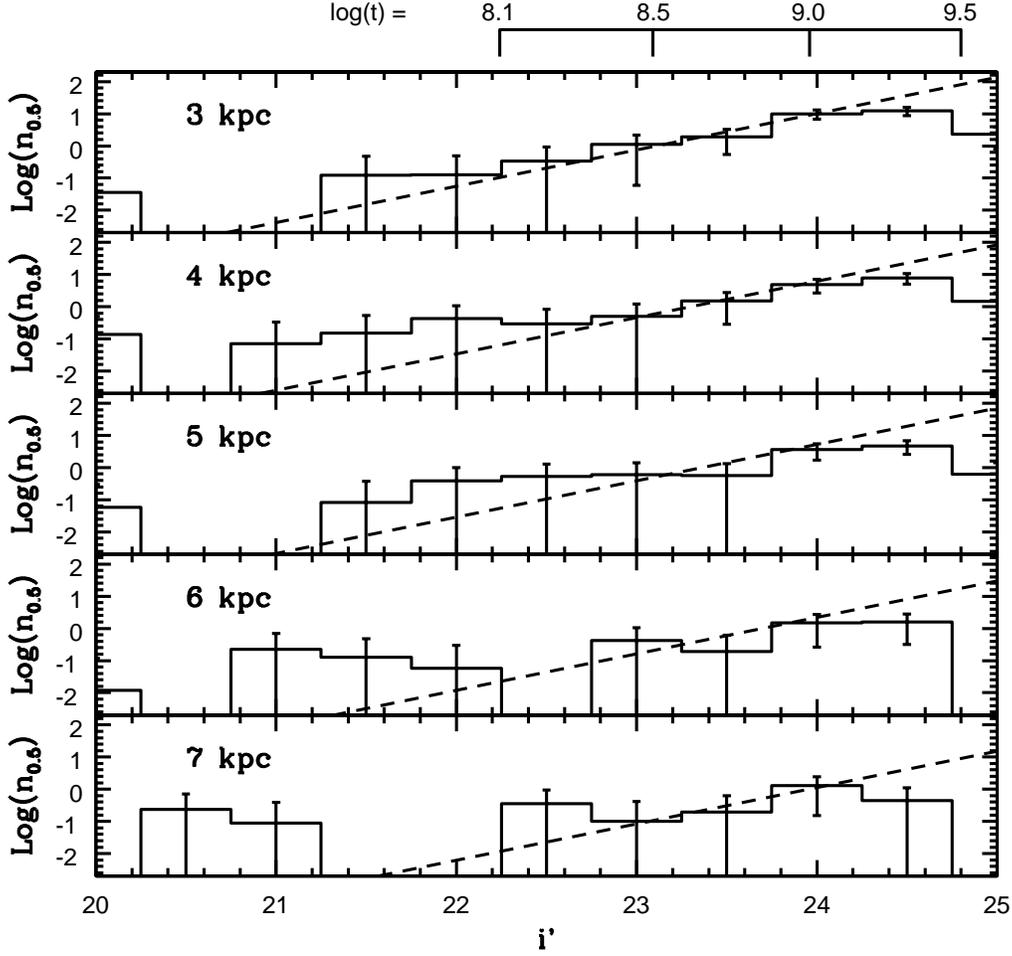}
\caption
{The $i'$ LFs of the extraplanar regions of M82. n$_{0.5}$ is the number of stars with 
$r'-i'$ between 0 and 2 per arcmin$^{2}$ per 0.5 magnitude $i'$ interval. The LFs have been 
corrected for contamination from foreground stars and background galaxies using 
number counts in the control fields. The dashed line shows the reference SF 
relation from Figure 7, shifted to match the number of stars with $i' =$ 23, 23.5, and 24 
at each D$_Z$. The peak AGB brightnesses in the Girardi et al. (2004) 
Z = 0.008 models are shown at the top of the figure, and this calibration suggests that 
significant numbers of AGB stars with ages log(t) $\sim 9$ are detected out to 
D$_Z = 7$ kpc.}
\end{figure}

\clearpage
\begin{figure}
\figurenum{15}
\epsscale{0.85}
\plotone{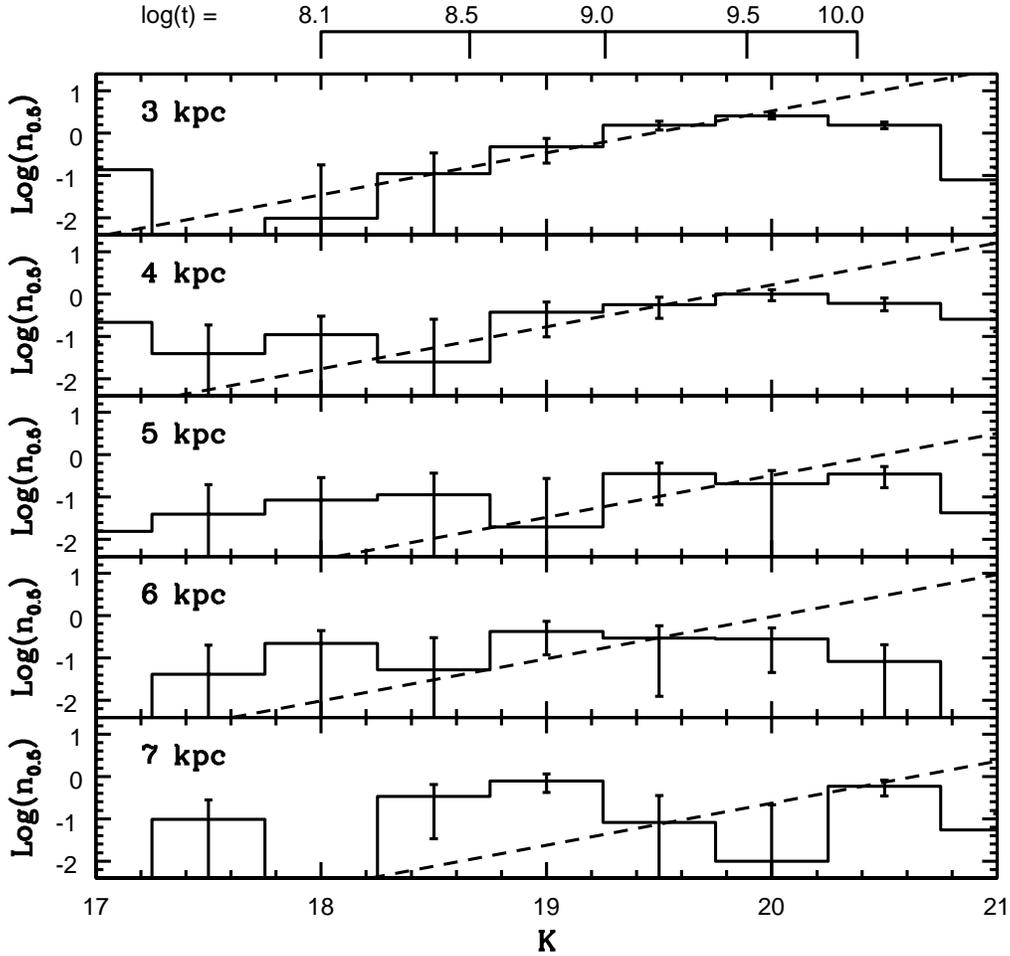}
\caption
{The same as Figure 14, but showing the $K$ LFs of stars with $H-K$ between 0 and 1 in the 
extraplanar regions of M82. The dashed line shows the reference 
SF relation from Figure 8 but shifted to match the number of 
stars with $K =$ 19, 19.5, and 20 at each D$_Z$.
The peak AGB brightnesses in the Girardi et al. (2002) 
Z = 0.008 models are shown at the top of the figure.}
\end{figure}

\clearpage
\begin{figure}
\figurenum{16}
\epsscale{0.85}
\plotone{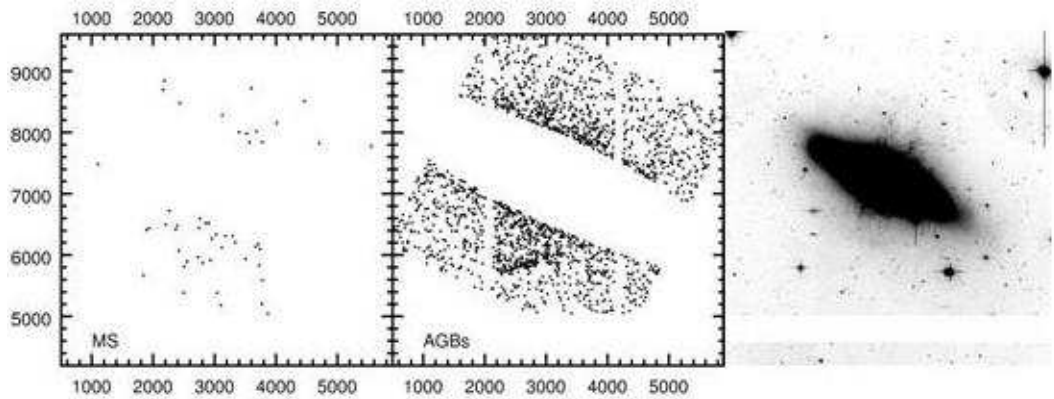}
\caption
{The spatial distribution of extraplanar objects with photometric properties that 
are consistent with either main sequence (MS) stars or AGB$+$RSG stars. 
Each panel covers $16 \times 16$ arcmin$^2$, and 
the corresponding section of the MegaCam $i'$ image is also shown. 
The co-ordinates are in pixel units. Note that 
the MS and AGB stars are concentrated along the minor axis 
of the galaxy, and that there is a tendency for the number densities of 
both types of objects to drop with increasing distance from the disk. The density of 
MS and AGB stars is higher to the south of the disk than to the north. 
M82 South is the compact collection of AGB stars to the south of the 
main body of M82, at (x,y) = (2600,6000).} 
\end{figure}

\clearpage
\begin{figure}
\figurenum{17}
\epsscale{0.85}
\plotone{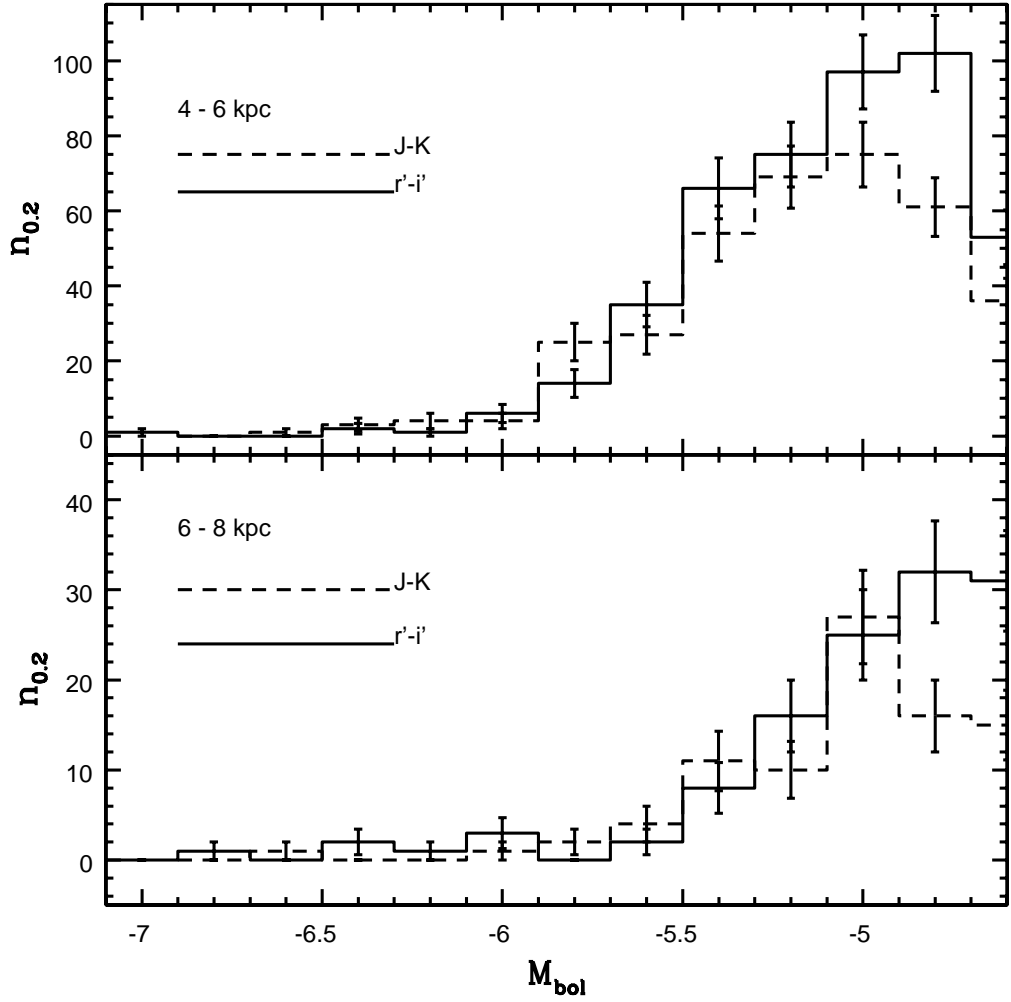}
\caption
{The bolometric LFs of objects with log(T$_{eff}$) between 3.54 
and 3.60 constructed from the $(i', r'-i')$ (solid line) and $(K, J-K)$ (dashed line) 
CMDs. n$_{0.2}$ is the number of stars per 0.2 mag interval in M$_{bol}$. The good agreement 
between the LFs constructed from the MegaCam and WIRCam data suggests that blending 
does not affect the bright AGB stellar content in the MegaCam images.}
\end{figure}

\end{document}